\documentclass[11pt,a4paper]{article}
\usepackage[margin=25mm]{geometry}
\usepackage[T1]{fontenc}
\usepackage{lmodern}
\usepackage{amsmath,amssymb,graphicx,booktabs,longtable,array,microtype}
\usepackage{caption,xurl}
\usepackage[colorlinks=true,linkcolor=black,citecolor=black,urlcolor=black]{hyperref}
\graphicspath{{figures/}}
\hypersetup{pdftitle={QICOM: Open-Source Software for Reproducible Image Quality Evaluation, Blinded Assessment, and Radiomic Comparison},pdfauthor={Fabio Mattiussi; Switinder Ghotra}}
\title{QICOM: Open-Source Software for Reproducible Image Quality Evaluation, Blinded Assessment, and Radiomic Comparison}
\author{Fabio Mattiussi\textsuperscript{1} and Switinder Ghotra\textsuperscript{2,3}\\[1.2ex]
\parbox{0.92\linewidth}{\centering\small
\textsuperscript{1}Integrated Diagnostics Institute of Southern Switzerland (IDISI), Ente
Ospedaliero Cantonale (EOC), 6900 Lugano, Switzerland\\[0.4ex]
\textsuperscript{2}School of Health Sciences (HESAV), University of Applied Sciences and Arts
Western Switzerland (HES-SO), 1011 Lausanne, Switzerland\\[0.4ex]
\textsuperscript{3}Department of Radiology, Lausanne University Hospital (CHUV) and University
of Lausanne (UNIL), Lausanne, Switzerland}}
\date{October 2026}

\begin{document}
\maketitle
\begin{abstract}
QICOM (Quality Image COMparator) is open-source software for quantitative image
comparison, paired radiomics, and blinded assessment. It saves processing settings, image
identities, measurements, and response mappings so a comparison can be reopened
and recalculated. We evaluate reference agreement, responses to controlled
perturbations, and recovery of saved analyses. Synthetic experiments cover
37 quantitative descriptors using independent numerical references
and separate upstream integration comparisons. The QICOM/MIRP backend meets the
reference tolerances for all 482 numeric IBSI~1 digital-phantom entries and all
1,622 entries across configurations A--E of the public CT reference case;
these comprise 1,804 feature/aggregation comparisons and 300 preprocessing diagnostics.
Restricting masked SSIM and HFEN
to complete kernel support removes the observed dependence on excluded pixels.
In a synthetic filtering example, stronger smoothing increases SSIM while worsening
MSE and HFEN relative to milder smoothing and changing GLCM contrast more strongly.
Reopened radiomics projects reproduce all 43 feature pairs per comparison, and
two blinded deliveries return four correctly mapped scripted responses. These
experiments document the calculations and data transfers linking the three workflows.

\end{abstract}

\section{Introduction}
The evaluation of medical imaging technologies, image-processing methods, and acquisition
protocols often combines quantitative analyses with qualitative assessments. Quantitative
approaches characterize image properties such as intensity differences, noise, tissue
contrast, and structural similarity. Region-of-interest (ROI) measurements include
signal-to-noise ratio (SNR) and contrast-to-noise ratio (CNR), whose interpretation requires
appropriate noise estimation, particularly when reconstruction or post-processing changes
noise characteristics. Reference-based metrics provide complementary measures of image
fidelity. For example, the ErisNet CT-denoising study combined mean squared error (MSE), peak
signal-to-noise ratio (PSNR), structural similarity index (SSIM), visual information fidelity
(VIF), and edge preservation measures with ROI measurements of attenuation and
noise~\cite{erisnet}. Azour et al.~\cite{azour} similarly combined SSIM and Fr\'echet
Inception Distance with radiologist assessments of image quality and lesion conspicuity.

Structured visual assessment examines anatomical visibility, perceived noise, artifacts, and
confidence in image interpretation. Reader scores can be evaluated using visual grading
analysis (VGA) and visual grading characteristics (VGC), alongside statistics that quantify
agreement between observers~\cite{precht}. These methods have been applied to compare
optimized X-ray protocols~\cite{sadosreis} and AI-reconstructed knee MRI with standard
imaging~\cite{ghotra_knee}, and to evaluate MRI protocols for total hip arthroplasty
planning~\cite{ghotra_hip}. When the objective concerns pathology detection, diagnostic
accuracy studies compare reader decisions against a predefined reference standard.
Sensitivity, specificity, and receiver operating characteristic (ROC) analysis characterize
detection performance, while free-response ROC (FROC) methods additionally incorporate lesion
localization. Analyses involving multiple readers and cases must account for both sources of
variability and the dependence between observations~\cite{obuchowski,chakraborty}. These
task-specific evaluations address whether changes in image appearance affect diagnostic
performance.

Radiomic comparison provides another perspective by quantifying changes in intensity
distributions, texture, and other image-derived features. The Image Biomarker
Standardization Initiative (IBSI) supplies standardized feature definitions and reference
data to support consistent extraction~\cite{ibsi}. Evaluating radiomic features alongside
image-quality measurements can help characterize processing-related changes that may affect
subsequent quantitative analyses.

Combining these evaluations requires numerical measurements, radiomic features, and reader
responses to remain linked to the same images and processing conditions. However, the
necessary methods are often distributed across separate applications and custom scripts.
Researchers may need to learn different interfaces, adapt images and masks to tool-specific
requirements, and reconcile exported results. A reproducible workflow must therefore preserve
image identifiers, masks, analysis settings, and blinded response mappings. Relevant settings
include slice or volume selection, intensity scaling, discretization, and the spatial support
of ROI calculations. Recording these choices supports repeated calculation and examination of
the relationship between numerical and observer-based outcomes.

Several open-source tools provide components of this workflow. scikit-image supplies general
image-processing algorithms~\cite{skimage}; PyTorch Image Quality (PIQ) implements
image-quality metrics~\cite{piq}; PyRadiomics provides configurable radiomic feature
extraction~\cite{pyradiomics}; 3D Slicer supports extensible medical-image
computing~\cite{slicer}; and ViewDEX supports blinded observer assessment of DICOM
images~\cite{viewdex}. Building on established computational libraries, QICOM (Quality Image
COMparator) brings quantitative image evaluation and comparison, blinded assessment, and
paired radiomics into a single application. Saved configurations, native projects, and
response mappings preserve the connections between inputs, processing conditions, and
results.

This study presents QICOM and evaluates three questions: whether its measurements agree with
analytical and external references; how they respond to controlled perturbations and spatial
restrictions; and whether saved inputs, settings, and outputs support re-execution of a
complete comparison. Numerical phantom experiments assess reference agreement and metric
behavior. IBSI digital-phantom and CT comparisons evaluate radiomic extraction, including
five preprocessing configurations applied to the same acquired CT case. A synthetic filtering
example follows the same images through all three workbenches, including export, reopening,
and repeated calculation, while scripted responses test blinded delivery and recovery of
source conditions. These experiments assess computational agreement and workflow
reproducibility; human perceptual validity and diagnostic performance require subsequent
evaluation.

\section{Workflow design}
\begin{figure}[htbp]
\centering
\includegraphics[width=\linewidth]{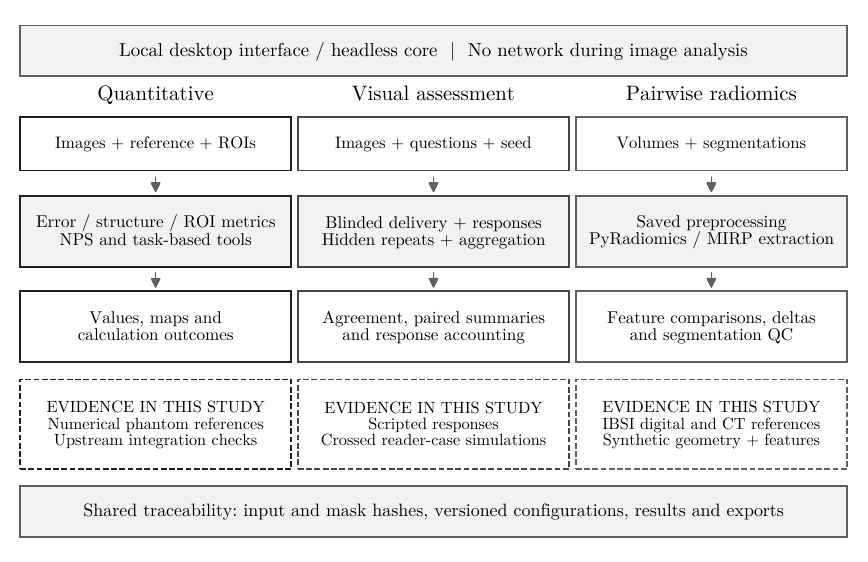}
\caption{QICOM's three workbenches, their inputs and outputs, and the numerical
and simulated-response experiments used in this study.}
\label{fig:architecture}
\end{figure}

\begin{figure}[htbp]
\centering
\includegraphics[width=\linewidth]{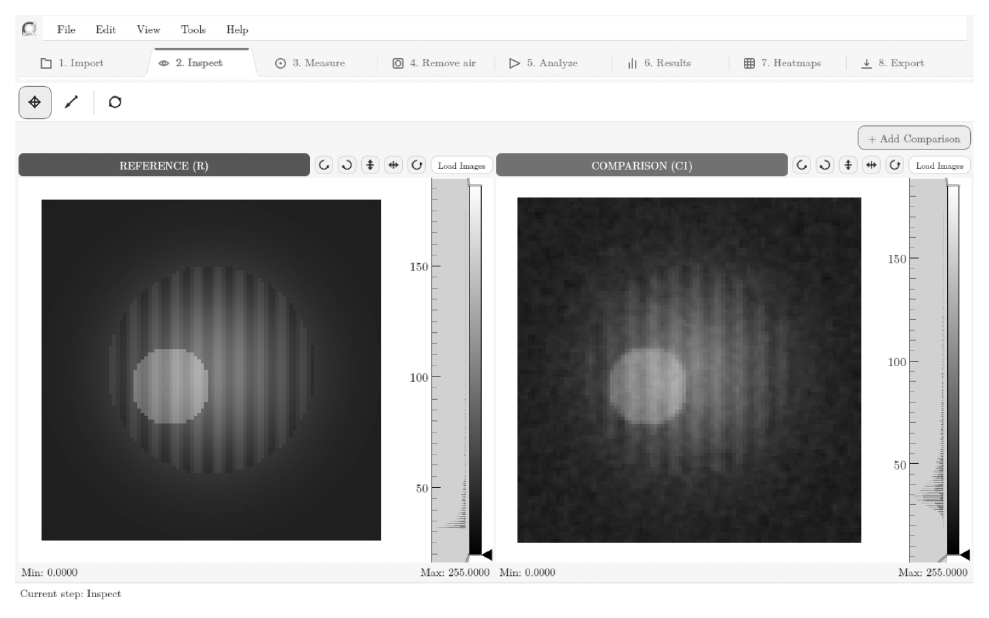}
\caption{The quantitative workbench captured from the running application with the
central-plane previews of the synthetic filtering inputs. The synchronized image
views show the reference and filtered image. Additional workbench captures appear in Supplementary
Section~\ref{app:gui}.}
\label{fig:gui}
\end{figure}

QICOM is Python-based desktop software for image evaluation. A Qt interface provides three
workbenches (Figures~\ref{fig:architecture} and \ref{fig:gui}). The quantitative workbench contains
37 descriptors covering intensity errors, structure, distributions, perceptual
backends, and ROI statistics. The radiomics workbench compares paired volumes
using masks and preprocessing settings. The qualitative workbench creates
blinded deliveries, collects structured responses, and aggregates returned data.
QICOM versions after the evaluated snapshots add a fourth section of Utility Tools
for semi-automatic segmentation, mask comparison, conversion, resampling, image
arithmetic, controlled degradation, DICOM header inspection and editing, folder
organization, decompression, and de-identification. Supplementary Section~\ref{app:tools} describes
these tools and the current workbenches; they were not part of the experiments
reported here.

The numerical, I/O, session, radiomics, and reporting packages run without Qt,
so the interface and headless experiments call the same computational functions.
Configurations record data range, ROI, geometry, discretization, and slice
aggregation. Quantitative analyses save input, configuration, and result hashes.
The outcome registry records computed values and unavailable, inapplicable,
invalid, failed, or cancelled calculations.

Native radiomics projects store arrays, masks, geometry, configuration, and
feature tables. Blinded deliveries use presentation identifiers and a separate
author-side mapping to recover source conditions from responses. Analysis runs
locally without network access. Optional learned metrics use previously installed
checkpoints with recorded identities. Scientific backends and supporting packages
are attributed in Supplementary Section~\ref{app:software}.

\section{Evaluation design}
\subsection{Reference agreement and controlled perturbation}
Quantitative phantoms comprise identity pairs, offsets, gains, structured patterns,
and realized noise arrays. For a reference $R$ and comparison $C=R+b$, without
clipping, the expected MSE is $b^2$ and signed bias is $b$. Independent SSIM~\cite{ssim}
and HFEN~\cite{hfen} calculations match the window, boundary, and kernel conventions
of the tested implementation. Direct PIQ and LPIPS calls provide separate
integration comparisons with matched inputs and parameters.

Graded noise, blur, and offsets probe metric responses. Masking fixtures modify
excluded pixels while keeping the included pixels fixed. They compare radial
erosion with restriction to complete kernel support. Spectral observer fixtures
use flat and colored noise and a nonconstant eye filter to compare prewhitening,
non-prewhitening, and eye-filtered calculations.

Radiomics uses the official IBSI~1 digital phantom and the public CT reference
case in NIfTI format~\cite{ibsi,lambin2017,ctphantom}. We evaluate all numeric
reference/tolerance rows in the digital-phantom worksheet and CT configurations
A--E. The optional MIRP backend supplies the additional feature families and
aggregation modes~\cite{mirp}; the original PyRadiomics subset is evaluated
separately. Acceptance follows the workbook's
magnitude-dependent truncation and row-specific absolute tolerances. The full
rule, near-zero safeguard, mapping, and quantitative tolerance classes are given
in Supplementary Section~\ref{app:methods}, including
Section~\ref{sec:ibsi-rule}. Workbook rows identify feature/aggregation
combinations; repeated aggregations of a feature occupy separate rows.

The qualitative experiments use synthetic responses and independent coefficient
calculations. Separate fixtures exercise blinding, hidden repeats, response
transport, and missingness accounting. Monte Carlo experiments estimate interval
coverage in a crossed reader--case model. Each generated dataset is one simulation
replicate. Assertions, transformed images, and repeated extractions are counted
within their respective experiments.

\subsection{Filtering comparison across the three workflows}
\label{sec:case-method}
The filtering demonstration uses an analytic float32 volume containing $16\times96\times96$ voxels with spacing $(1,1,2)$\,mm in
physical $(x,y,z)$ order. A smooth background contains a circular intensity
inclusion and a sinusoidal pattern. Gaussian noise with standard deviation 8 and
seed 260910 is followed by in-plane Gaussian filtering with standard deviations
0.65 and 1.5 pixels. The script saves its configuration before computation and
writes clean, noisy, and filtered NIfTI images.

MSE, SSIM, and HFEN compare the central plane with the clean reference at data
range 255. Radiomics uses the same volumes and a fixed interior ROI of 38,520
voxels. Extraction includes original-image first-order and GLCM features, no
intensity normalization, bin width 8, ROI-minimum binning origin, and target
spacing equal to source spacing. Quantitative exports and native radiomics
projects are reopened, and calculations are repeated from the saved data.

Two blinded deliveries present the clean reference and both filtered conditions.
Their central-plane previews are rounded and clipped to 8 bits for display;
measurements use the reloaded float32 arrays. Four arbitrary scripted responses
exercise export, import, and recovery of source labels. The response values are
transport markers, excluded from the image-quality outcomes.

\subsection{Experimental records}
The companion files contain inputs, settings, expected and observed values,
row-level statuses, and source hashes. Each experiment identifies its software
snapshot and input files. Supplementary Section~\ref{app:methods} specifies the
formulas and tolerances; Section~\ref{app:reproduction} identifies the source and
dependency records. Figures are drawn from these records. GUI images are native
Qt captures using the saved synthetic comparison. Figure~\ref{fig:ct-reference}
shows the CT input and its supplied mask.

\section{Results}
\subsection{Agreement with numerical and external references}
\label{sec:agreement}
Quantitative outputs agreed with the analytic and independently implemented
references within the experiment tolerances. Maximum absolute discrepancies were
$1.67\times10^{-8}$ for MSE, $4.63\times10^{-7}$ for SSIM, and $1.80\times10^{-7}$ for HFEN.
The direct PIQ and LPIPS comparisons also met their integration tolerances.

\begin{figure}[htbp]\centering
\includegraphics[width=\linewidth]{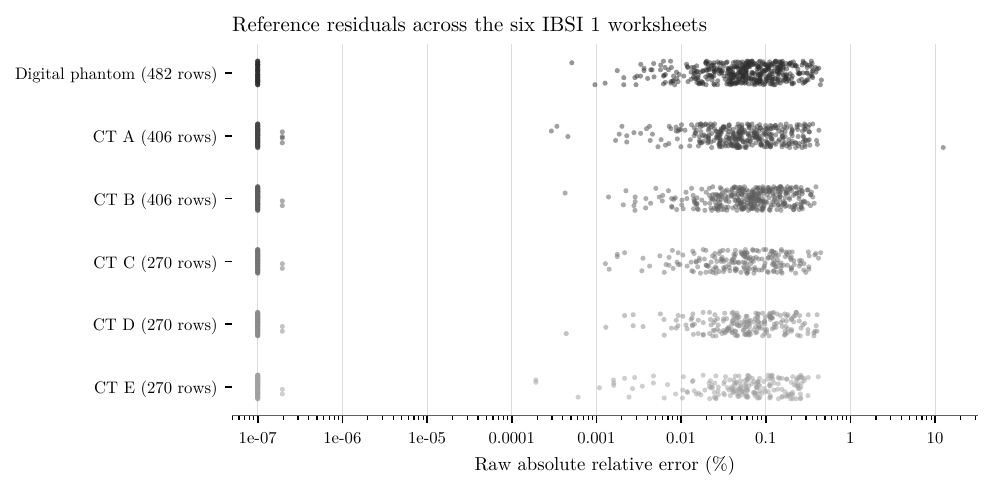}
\caption{Raw absolute relative errors against IBSI reference values. Each point represents one worksheet row with a nonzero reference; zero residuals are placed at the left plotting floor. Acceptance uses the row-specific absolute tolerance and truncation rule in the Methods.}
\label{fig:ibsi}
\end{figure}

The QICOM/MIRP backend met the reference tolerances for all 482 digital-phantom
entries and all 1,622 CT entries across configurations A--E
(Table~\ref{tab:ibsi}). All 1,804 feature/aggregation comparisons and all 300
preprocessing diagnostics passed. Every numeric reference/tolerance row in the
six worksheets was evaluated. Figure~\ref{fig:ibsi} shows the residuals.

Of the 482 digital-phantom entries, 457 (94.8\%)
had a tabulated absolute tolerance of zero; the remaining 25
had positive tolerances.
The largest raw relative error was 12.49\% for volume
fraction at 90\% intensity (IVH V90) in CT A: reference
$6.98\times10^{-5}$, raw absolute error
$8.72\times10^{-6}$, and absolute tolerance
$1.03\times10^{-5}$.
\begin{table}[htbp]
\centering\small
\begin{tabular}{lrrr}
\toprule
IBSI 1 configuration & Features & Diagnostics & Passed / total \\
\midrule
Digital phantom & 482 & 0 & 482 / 482 \\
CT A & 346 & 60 & 406 / 406 \\
CT B & 346 & 60 & 406 / 406 \\
CT C & 210 & 60 & 270 / 270 \\
CT D & 210 & 60 & 270 / 270 \\
CT E & 210 & 60 & 270 / 270 \\
\bottomrule
\end{tabular}
\caption{Reference agreement through the QICOM/MIRP backend. The five CT configurations use one public case; entries count feature/aggregation combinations and preprocessing diagnostics.}
\label{tab:ibsi}
\end{table}

The separate PyRadiomics experiment met the tolerances for its 104 mapped
digital-phantom entries (104/482 worksheet rows, 21.6\%). Its largest raw relative
discrepancy was 0.6992\% for minor-axis length: 9.24490 versus 9.31\,mm. The raw
absolute error was 0.065095\,mm; the truncated comparison error and tolerance
were both 0.06\,mm.
\begin{figure}[htbp]\centering
\includegraphics[width=\linewidth]{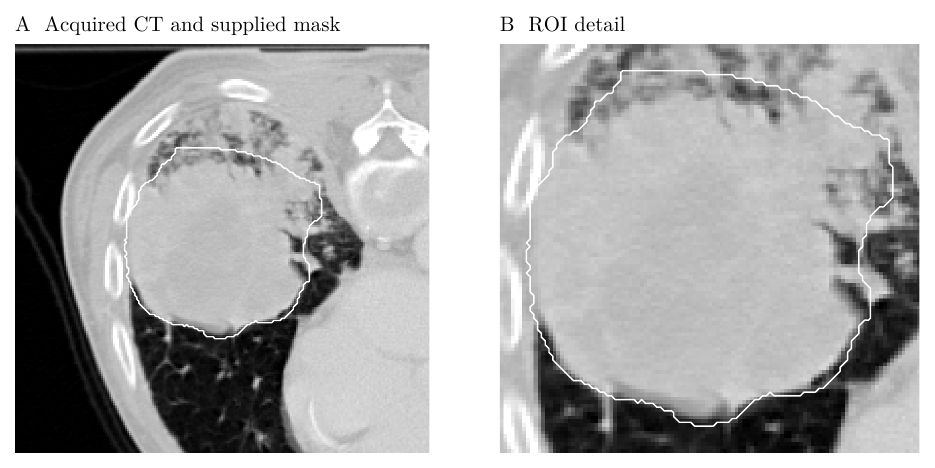}
\caption{Public IBSI CT reference case in NIfTI format. Axial plane through the largest mask cross-section, displayed at $[-1000,400]$ HU; the white contour is the supplied mask. The inset enlarges the same ROI. Display adapted from the CT image and mask supplied by Lambin, CancerData (2016)~\cite{ctphantom,lambin2017}. The CT-derived image and contour retain the source license, \href{https://creativecommons.org/licenses/by-nc/3.0/}{CC BY-NC 3.0}.}
\label{fig:ct-reference}
\end{figure}

\subsection{Behavior under perturbation and spatial restriction}
\label{sec:perturbation}
\begin{figure}[htbp]\centering
\includegraphics[width=\linewidth]{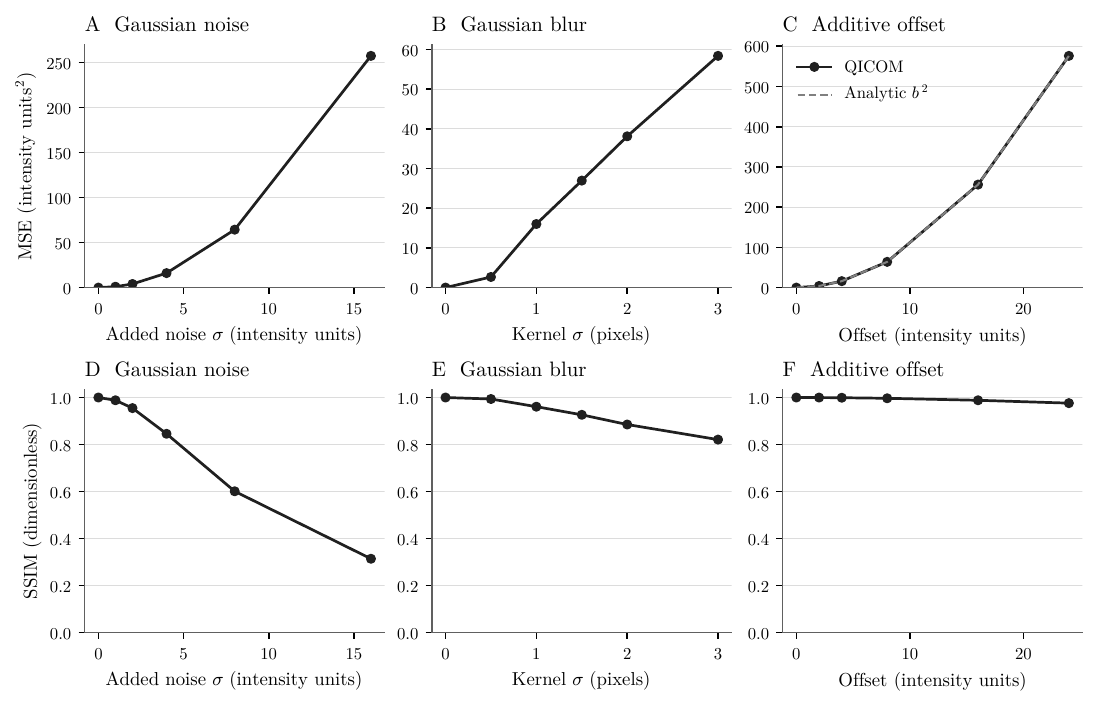}
\caption{MSE and SSIM responses to noise, Gaussian blur, and intensity offsets on the synthetic phantom. Points are computed values; lines connect successive levels.}
\label{fig:degradation}
\end{figure}

Figure~\ref{fig:degradation} shows the responses to noise, blur, and intensity
offsets. MSE increased across the tested levels of all three perturbations.
SSIM decreased with noise and blur and remained closer to unity under offsets.

With radial erosion, masked SSIM changed when excluded comparison pixels changed.
With complete kernel support, masked SSIM and HFEN were invariant to the tested
excluded-pixel perturbations. With radial erosion, the SSIM values were 0.842831
and 0.800035 for the two exterior-pixel settings.

PW and NPW returned equal values for the flat-spectrum fixture and different
values for the colored spectra. NPWE with the nonconstant eye filter differed
from both in all three fixtures (Table~\ref{tab:detectability}).
\begin{table}[htbp]
\centering\small
\begin{tabular}{lrrr}
\toprule
Spectrum / task & $d'_{PW}$ & $d'_{NPW}$ & $d'_{NPWE}$ \\
\midrule
flat & 1.500000 & 1.500000 & 1.397820 \\
colored flat task & 1.517647 & 1.279204 & 1.142984 \\
colored gaussian task & 0.987565 & 0.767542 & 0.507999 \\
\bottomrule
\end{tabular}
\caption{Observer outputs for synthetic spectrum and task fixtures.}
\label{tab:detectability}
\end{table}

\subsection{Simulated reader responses}
Ordinal agreement coefficients matched their independent references in the
applicable simulated fixtures. All 127 scripted response rows were recovered
under their source conditions. None of the 96 hidden-repeat pairs violated the
specified spacing. Empirical bootstrap interval coverage was 95/100 datasets in
both the complete-data and 15\% missingness settings (Table~\ref{tab:simulation}).
\begin{table}[htbp]
\centering\small
\begin{tabular}{rrrr}
\toprule
Missing probability & Coverage & MC standard error & Mean CI width \\
\midrule
0.00 & 0.95 & 0.0218 & 0.943 \\
0.15 & 0.95 & 0.0218 & 1.016 \\
\bottomrule
\end{tabular}
\caption{Crossed reader--case simulations: 100 generated datasets per setting. The 95\% Wilson interval for coverage is approximately 0.888--0.978 in both settings.}
\label{tab:simulation}
\end{table}

\subsection{A complete synthetic filtering comparison}
\label{sec:case}
\begin{figure}[htbp]\centering
\includegraphics[width=\linewidth]{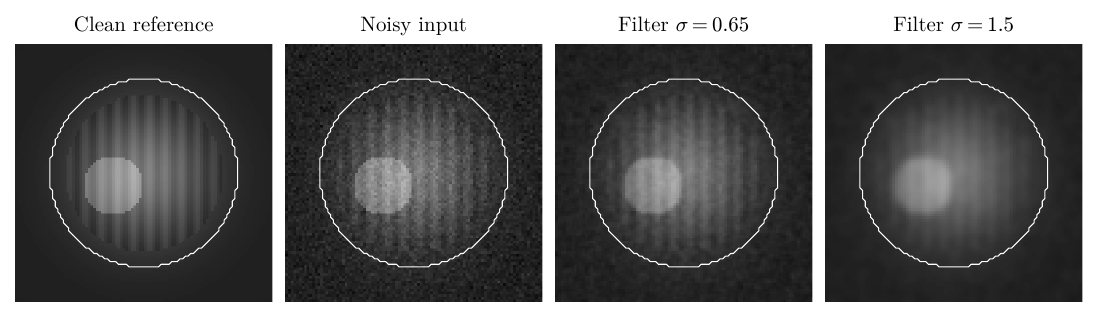}
\caption{Central plane of the clean, noisy, and filtered volumes. The common display range is 0--255; the white contour marks the radiomic ROI.}
\label{fig:filtering}
\end{figure}

Both filters reduced MSE relative to the noisy input
(Figure~\ref{fig:filtering}; Table~\ref{tab:filtering}). Increasing Gaussian smoothing
from 0.65 to 1.5 pixels increased SSIM, MSE, and HFEN.
\begin{table}[htbp]
\centering\small
\begin{tabular}{lrrrr}
\toprule
Condition & MSE $\downarrow$ & SSIM $\uparrow$ & HFEN $\downarrow$ & GLCM contrast \\
\midrule
Noisy & 63.001 & 0.6198 & 0.5932 & 3.4814 \\
$\sigma=0.65$ & 15.849 & 0.8724 & 0.5054 & 1.3488 \\
$\sigma=1.5$ & 18.421 & 0.9119 & 0.6998 & 0.6356 \\
\bottomrule
\end{tabular}
\caption{Quantitative and radiomic results for the filtering example. Clean-reference GLCM contrast is 1.4411. Arrows indicate decreasing error or increasing similarity. MSE is in squared arbitrary intensity units; the other displayed measures are dimensionless.}
\label{tab:filtering}
\end{table}

GLCM contrast differed from
the clean reference by +141.6\% for noise alone,
-6.4\% after mild filtering, and -55.9\% after stronger
filtering. In contrast, mean ROI intensity changed by less than 0.085\%
across all three conditions.

All saved quantitative values were recovered on JSON reload, and repeated calls
returned matching input, configuration, and result hashes. Reopening each of the
three native radiomics projects preserved arrays, masks, settings, and feature
tables; re-extraction reproduced all 43 feature pairs
exactly in the recorded environment. The two blinded deliveries transported
4 scripted answer records, all restored to the
correct source condition.

\section{Discussion}
\subsection{Filtering changes intensity, structure, and texture differently}
Stronger filtering increased SSIM while increasing MSE and HFEN relative to
milder filtering. It also moved GLCM contrast further from the clean reference,
despite a small change in mean ROI intensity. An SSIM-only choice would therefore
favor a condition with greater intensity and high-frequency error in this
example. The distinction between error and structural similarity has been studied
previously~\cite{hore}. Here, measuring texture alongside those scores exposes
another consequence of smoothing that matters when processed images will be used
for radiomic analysis.

QICOM keeps those measurements linked to the images that produced them. The same
comparison can be reopened, its features recalculated, and its images prepared
for blinded assessment. A subsequent reader study could determine whether the
measured tradeoff affects a specified visual task.

\subsection{Spatial support and model definitions}
With radial mask erosion, pixels outside the ROI still entered local calculations
through corners of the square convolution window. Restricting centers to complete
kernel support removed the observed dependence. Thus an ROI definition must
specify both the pixels whose scores are averaged and the neighboring pixels
used to calculate those scores.

The observer fixtures expose a similar dependence on model definition. Flat-noise
inputs can give coincident PW and NPW values, while colored spectra separate their
weighting. Multiple spectra and a nonconstant eye filter distinguish the
models that coincide under flat noise. Both examples favor saving kernel
support, normalization, geometry, and model parameters with the output.

No-reference scores depend on their reference model in the same way. The NSS
distances evaluated here are defined relative to QICOM's declared distribution,
and the distributed NIQE model is fitted on photographs. For CT, QICOM versions
after the evaluated snapshot add RIQE, a NIQE-style model fitted on full-dose CT
with a declared Hounsfield-unit mapping and body masks, archived with its code,
fitting manifest, and validation~\cite{riqe}. QICOM computes it only for images
in Hounsfield units and scores a processed image on the masks of its source. In
its validation, RIQE ranked degraded images as worse than their source, but its
preferences among filtered images did not follow lesion signal; it is therefore
not a criterion for choosing a denoiser. RIQE was not part of the experiments
reported here.

\subsection{Contribution relative to existing tools}
A scripted combination of scikit-image, PIQ, and PyRadiomics provides flexible
access to the numerical methods. An extensible platform such as 3D Slicer supports
broader imaging workflows. QICOM concentrates on paired comparison and blinded
delivery, with saved configurations, native projects, and source-label recovery.
Maintaining these connections also requires adapters and versioned persistent
formats as the underlying libraries evolve.

Analytic references, upstream comparisons, and external phantoms play different
roles in evaluating this arrangement. Independent calculations challenge formulas
and conventions; same-backend comparisons check data and parameter handling;
IBSI tests agreement with external reference values. The small residuals reported
here measure computational agreement, rather than the size of an image-quality
improvement.

The complete digital-phantom mapping adds local intensity, histogram, volume
histogram, distance-zone features, and the specified texture aggregations.
The CT comparison also exercises interpolation, resegmentation, and discretization
on acquired data. Agreement therefore covers the QICOM adapter and MIRP extraction
under each reference preset. PyRadiomics retains its own, narrower mapping.
Backend identity is saved with each extraction because implementations and
aggregation conventions determine which comparisons are meaningful. Exact spatial
sums for Moran's $I$ and Geary's $C$ avoid sampling variability in these two outputs.

The QIBA metrology framework distinguishes bias, repeatability, and reproducibility
in quantitative imaging~\cite{qiba}. This study addresses reference agreement and
re-execution under controlled technical conditions. Repeatability across
acquisitions and reproducibility across scanners or sites require additional
experiments with physical phantoms and acquisition variability, with uncertainty
estimated at the corresponding experimental unit.

\subsection{Scope and limitations}
\label{sec:limits}
\label{app:component-limits}
The study uses synthetic images, one public CT case, scripted responses, and one
recorded software environment. Configurations A--E are five analyses of that same
CT case. The complete worksheet coverage concerns IBSI~1 and its specified
parameters. Human perceptual validity, clinical utility,
cross-site performance, and comparative usability remain untested. QICOM is not
a medical device. The filtering example is a single retrospective demonstration.

The NSS distances depend on their reference distribution. Their score scales, the composite
index's normalization anchors and equal weights, and the one-dimensional observer
model lack human calibration. Edge experiments cover axis-aligned noiseless
Gaussian profiles. Blinding removes tested text and identifying EXIF fields;
burned-in labels and arbitrary content in retained ICC profiles remain outside
that metadata check. Response completeness requires an assignment roster to
count assignments with no returned record. The external radiomic comparison covers
the six IBSI~1 worksheets; IBSI~2 filtering and vendor-specific image encodings
require separate evaluation.

\subsection{A subsequent observer study}
A reader study should specify the visual task before selecting images and
endpoints, for example preservation of low-contrast structure after denoising.
Processing conditions, eligibility criteria, display and windowing settings,
training, and randomization should be fixed before data collection. The companion
feasibility protocol proposes eight readers, 24 base cases, four hidden repeats,
and two training presentations. These planning numbers require a pilot before
a confirmatory sample size can be calculated.

Ordinal inter-reader agreement, within-reader repeat consistency, and the paired
condition effect address different aspects of the assessment. Analysis should
retain crossed reader and case dependence and define the handling of skipped,
not-assessable, and entirely absent responses. An assignment roster supplies the
denominator for response completeness. Independent replication and a controlled
usability comparison would then evaluate study execution as well as calculations.

\section{Availability and reproducibility}
QICOM source is maintained at \url{https://github.com/Metallogik/QICOM} under the
Apache~2.0 license. The repository is currently private and will be made public
before arXiv submission. The evaluated development snapshots, dependency versions,
source hashes, and experiment records are listed in Supplementary
Section~\ref{app:reproduction}. The original experiments and the IBSI extension
have separate source archives. MIRP is an optional dependency distributed under
EUPL~1.2; the public CT data are available under CC~BY-NC~3.0 from their original
source~\cite{ctphantom}. The RIQE model included in later QICOM versions is
archived under CC~BY~4.0 at \url{https://doi.org/10.5281/zenodo.23055559}.

The experiment bundle is held locally and contains the scripts, synthetic inputs,
results, and both source archives. It is separate from the software repository.
External IBSI resources and learned checkpoints are listed by source URL and hash
and must be obtained separately. Figure~\ref{fig:ct-reference} contains a display
adaptation of the CT data under their original CC~BY-NC~3.0 license.

\section{Conclusion}
QICOM connects numerical comparison, paired radiomics, and blinded delivery
through saved images, settings, and results. The experiments document reference
agreement and responses to controlled perturbations. Reopening the filtering
projects reproduced their measurements and recovered the scripted responses under
their source conditions. The next step is a task-specific reader study with
realistic acquisition variability.

\begingroup
\fontsize{9.5}{10.2}\selectfont
\let\originalbibliography\thebibliography
\renewcommand{\thebibliography}[1]{\originalbibliography{#1}\setlength{\itemsep}{0pt}\setlength{\parskip}{0pt}}
\bibliographystyle{unsrt}
\bibliography{references}
\endgroup

\clearpage
\appendix
\section{Detailed experimental methods}
\label{app:methods}
\subsection{Quantitative phantom experiments}
Error and ROI measures are tested using deterministic objects with known
intensity distributions, offsets, gains, and residuals. For a nonconstant reference
image $R$ and comparison $C=R+b$, without clipping or quantization,
\begin{equation}
 \mathrm{MSE}=b^2,\qquad \mathrm{MAE}=\mathrm{RMSE}=|b|,\qquad
 \mathrm{bias}=b,\qquad \mathrm{SD}(C-R)=0.
\end{equation}
For $b\ne0$, $\mathrm{PSNR}=10\log_{10}(L^2/b^2)$ with data range $L$.
Identity and undefined-input cases are handled separately. ROI references use
the actual sample values and variance conventions. Finite-realization variance
and population noise variance are calculated separately.

Statistical and spatial references include histogram and order-statistic
calculations, structured patterns, and independent local-window calculations.
SSIM conventions are matched to the tested implementation, including window,
covariance normalization, boundaries, and dynamic range~\cite{ssim}.
PSNR and SSIM comparisons follow their respective error and structural
definitions~\cite{hore}.

The structural metric catalog additionally includes MS-SSIM~\cite{msssim},
FSIM~\cite{fsim}, GMSD~\cite{gmsd}, VIF~\cite{vif}, and HaarPSI~\cite{haarpsi}.
The PIQ version and QICOM configuration identify the evaluated implementations.
NMI uses the ratio
$[H(R)+H(C)]/H(R,C)$~\cite{nmi}, estimated from discrete histogram masses.
Intensity Wasserstein-1 is a distance between one-dimensional intensity
distributions~\cite{wasserstein}.

For edge preservation, QICOM uses a high-pass correlation measure in the tradition
of Sattar et al.~\cite{epi}: $\mathrm{EPI}=\mathrm{corr}(\Delta R,\Delta C)$,
where $\Delta$ is the discrete Laplacian. This is distinct from the normalized
gradient-error expression labeled EPI in the ErisNet article~\cite{erisnet}.
HFEN follows the Laplacian-of-Gaussian error approach used in MRI
reconstruction~\cite{hfen}; QICOM reports
$\|\mathrm{LoG}_{\sigma}(C-R)\|_2/\|\mathrm{LoG}_{\sigma}(R)\|_2$ with
$\sigma=1.5$ pixels or voxels by default. Kernel support, dimensionality, and
boundaries are part of the tested convention. The Laplacian-variance sharpness
descriptor is a focus measure~\cite{blur}.

For structural and learned metrics, direct PIQ or LPIPS calls use matched inputs,
normalization, channels, parameters, and installed weights. These comparisons are
labeled integration checks. LPIPS and DISTS provide learned perceptual
distances~\cite{lpips,dists}. Graded degradations provide metric response curves.
The tested LPIPS and DISTS backbones are AlexNet~\cite{alexnet} and
VGG~\cite{vgg}, respectively; backbone and perceptual-metric weights are distinct
resources recorded in the checkpoint manifest.

The NSS feature distance compares a 36-dimensional image feature vector with a
reference distribution using Mahalanobis distance. The NSS patch-distribution
distance compares the mean patch features with that distribution using pooled
covariance. Their feature construction draws on BRISQUE~\cite{brisque} and
NIQE~\cite{niqe}. QICOM fits its default distribution to synthetic images with a
$1/f$ spectrum and accepts a user-fitted local distribution. The output names
identify these two distances; the identifiers \texttt{brisque} and \texttt{niqe}
are retained only for compatibility with saved projects. The former has no
human-score-trained regressor, and neither output reproduces the original score
scale. EPI is displayed as ``EPI (Laplacian correlation)''.

\begin{figure}[htbp]\centering
\includegraphics[width=\linewidth]{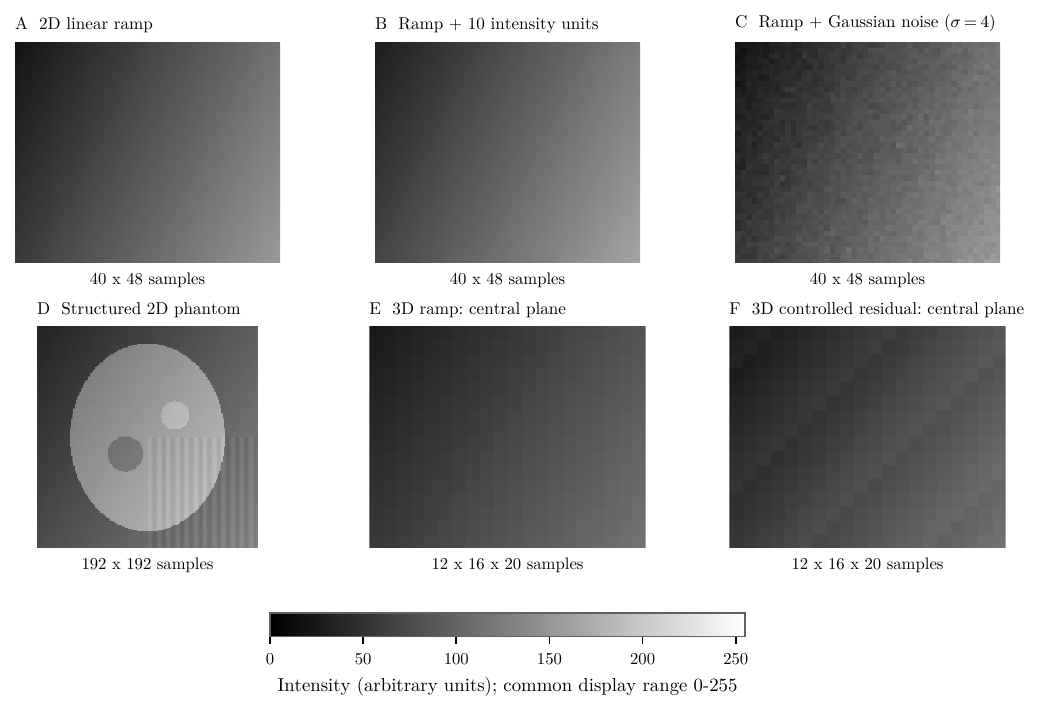}
\caption{Representative synthetic inputs generated by code, with known intensity and spatial transformations. Display normalization in this montage is only for visualization; computations use the saved numerical arrays.}
\label{fig:phantoms}
\end{figure}

The quantitative seed was 260909. Native arrays used shapes $40\times48$ and
$12\times16\times20$; structural comparisons used $192\times192$ images. Inputs were
float32 and the dynamic range was 255. The artifact stores 19 array objects
as NPZ and NIfTI, including identity copies, ramps, offsets, gains, controlled residuals,
and structured patterns. The object count includes derived and identity copies.
Figure~\ref{fig:phantoms} shows representative inputs.
Degradation series used Gaussian noise standard deviations $0,1,2,4,8,16$, blur standard
deviations $0,0.5,1,1.5,2,3$ pixels, and offsets $0,2,4,8,16,24$, producing
342 metric responses. Full settings and processing conventions are in the
executable configuration and reference code. The NSS experiment uses 96-pixel patches and
covariance ridge 0.01 and a fixed synthetic reference model for the QICOM variant.

Numerical acceptance uses $|o-e|\le a+r|e|$, where $o$ and $e$ are observed and expected
values. Table~\ref{tab:tolerances} states the main tolerance classes. Undefined or infinite
outputs are compared by their specified state. A separate noise-variance distribution check
uses a five-standard-error bound, separately from the float64 error tolerance.
\begin{table}[htbp]
\centering\small
\begin{tabular}{lrr}
\toprule
Class & Absolute $a$ & Relative $r$ \\
\midrule
float32 & 1e-06 & 1e-05 \\
float64 & 1e-10 & 1e-10 \\
hfen & 1e-05 & 1e-04 \\
integration & 2e-06 & 2e-05 \\
nss & 1e-07 & 1e-07 \\
ssim & 2e-05 & 2e-05 \\
\bottomrule
\end{tabular}
\caption{Numerical acceptance tolerances for the convention-matched quantitative references.}
\label{tab:tolerances}
\end{table}

\subsection{Expert diagnostics and image I/O}
Noise experiments test both numerical NPS normalization and expected behavior of
finite realizations. Parseval-based comparisons use the discrete window, detrending,
frequency-grid, and spacing conventions of the experiment. The NPS and edge-response
tools are interpreted within the established noise/transfer-function
framework~\cite{nps,solomon}. Known translations test phase-correlation alignment
diagnostics~\cite{guizar}; overlap and surface-distance quantities are interpreted
with voxel and physical-space conventions~\cite{taha}. Gaussian edges
and known geometries provide references for edge response, displacement, overlap,
and physical distances. Synthetic NIfTI and DICOM files test selected intensity
and geometry conversions.

Observer-model labels are evaluated independently of numerical integration. Let
$S(f)=|W(f)T(f)|^2$ denote the squared task spectrum after transfer and let $N(f)$
denote the NPS. With matched integration conventions, a prewhitening (PW) expression
and a non-prewhitening (NPW) expression are respectively
\begin{equation}
 d_{\mathrm{PW}}^2=\int \frac{S(f)}{N(f)}\,df,
 \qquad d_{\mathrm{NPW}}^2=\frac{\left[\int S(f)\,df\right]^2}
                              {\int S(f)N(f)\,df}.
\label{eq:observers}
\end{equation}
An eye-filtered NPW model additionally specifies an eye filter~\cite{solomon}.
Each numerical output is compared with the integral associated with its model label.

The diagnostic seed was 2026090913. NPS experiments used 128 independent $64\times64$
noise arrays with standard deviation 7 and spacings $(0.7,0.4)$ mm. White and filtered
noise were checked against the realized-array Parseval identity and finite-ensemble theory.
Mean and planar detrending, rectangular/Hann/Tukey windows, doubled spacing, and input
guards were included. Gaussian edges had standard deviations $0.8,1.5,2.5$ mm at 0.1 mm
spacing. The 10--90\% width and MTF50 were compared with the corresponding Gaussian
values, allowing 0.2\% and 2\% relative discretization error respectively. The MTF-curve
absolute tolerance was 0.02. Only axis-aligned, noiseless edges were tested; nonzero
edge angles are rejected by the implementation.

PW and NPW references used the same one-dimensional frequency grid and trapezoidal
integration. NPWE additionally used the illustrative filter
$E(f)=(f+0.01)\exp(-2f)$; its numerator is
$[\int S(f)E(f)^2\,df]^2$ and denominator is
$\int S(f)N(f)E(f)^4\,df$. Flat and colored noise spectra were evaluated with
this fixed filter and with a unity filter.

\subsection{Radiomic verification and derived quantities}
\label{sec:ibsi-rule}
The external comparison loads hash-verified images, masks, and the IBSI~1
reference workbook. Six manifests map every row having both a numeric reference
and an absolute tolerance: 482 in the digital phantom, 406 each in CT A and B,
and 270 each in CT C, D, and E. The original IBSI study standardized 169 of 174
investigated feature definitions~\cite{ibsi}; the worksheet rows additionally
separate aggregation modes and preprocessing diagnostics. The original
PyRadiomics manifest maps 104 digital-phantom rows and remains a separate experiment.

MIRP~2.7.0 supplies eleven feature families and the required 2D, 2.5D, and 3D
aggregation modes~\cite{mirp}. Presets use MIRP preprocessing after QICOM NIfTI
loading. Custom profiles instead use QICOM preprocessing before MIRP extraction.
The verifier evaluates presets. The CT image and supplied mask contain
$204\times201\times60$ voxels in physical $(x,y,z)$ order; the five configurations
operate on the same acquired case~\cite{lambin2017,ctphantom}.

The digital phantom uses its integer intensities without discretization or
resampling. CT A retains the acquired spacing; CT B interpolates in-plane to
2\,mm; CT C--E use isotropic 2\,mm sampling. Image interpolation is linear in B--D
and cubic in E. Mask interpolation is linear with inclusion threshold 0.5.
Anti-aliasing is disabled. Table~\ref{tab:ct-config} gives the intensity settings.
\begin{table}[htbp]
\centering\small
\begin{tabular}{llll}
\toprule
CT & Resegmentation (HU) & Texture bins & IVH bins \\
\midrule
A & $[-500,400]$ & Width 25 HU & None \\
B & $[-500,400]$ & Count 32 & None \\
C & $[-1000,400]$ & Width 25 HU & Width 2.5 HU \\
D & $\pm3$ SD & Count 32 & None \\
E & $[-1000,400]$ and $\pm3$ SD & Count 32 & Count 1000 \\
\bottomrule
\end{tabular}
\caption{IBSI CT preprocessing configurations. IVH: intensity--volume histogram. A--B use the specified planar aggregations; C--E use volumetric aggregations.}
\label{tab:ct-config}
\end{table}

QICOM calculates Moran's $I$ and Geary's $C$ from exact inverse-Euclidean-distance
sums over the intensity ROI. Linear FFT convolutions on the physical voxel grid
compute the sums with zero self-weight, replacing the MIRP sampling approximation
for those two outputs. Small 2D and 3D fixtures compare these results against
explicit all-pairs calculations, including anisotropic and sheared geometry.
Preprocessing diagnostics are calculated from the actual input, interpolated,
and resegmented arrays returned during extraction. Each receipt records backend,
version, effective settings, source hashes, and the provider of each output.

Each mapped entry uses the workbook's absolute tolerance $t$ in the feature's units.
For reference $e$, observation $o$, and raw difference $d=|o-e|$, define
\[
k=2-\lfloor\log_{10}|e|\rfloor,\quad s=10^k,\qquad
 d^*=\frac{\operatorname{trunc}[(d+\epsilon)s]}{s},\qquad
\mathrm{PASS}\ \Longleftrightarrow\ d^*\le t,
\]
where $|e|>10^{-12}$ and $\epsilon=\operatorname{ulp}(1)=2^{-52}$ is the
implementation's floating-point guard. Truncation retains the decimal resolution
of three significant digits of the reference, following the workbook's
\texttt{ROUNDDOWN} comparison. For $|e|\le10^{-12}$, QICOM instead compares
the untruncated $d$ with $t$: this near-zero safeguard avoids the
workbook fallback that could erase subunit errors. The rule is identical for all six worksheets; each reference/tolerance pair
$(e,t)$ is retained in the comparison CSV.

The CSV columns \texttt{absolute\_error}, \texttt{comparison\_error}, and
\texttt{absolute\_tolerance} record $d$, the error used for acceptance, and $t$,
respectively. The rule truncates the error at the reference precision rather than
rounding $e$ and $o$ separately. For $|e|>10^{-12}$, a tabulated tolerance of zero
requires $d^*=0$ and can therefore admit a nonzero raw difference $d$.

Three repeated extractions characterize same-environment computational repeatability.
A separate $16\times20\times24$ voxel object with intensity $10+2x+3y+5z$ and an
interior rectangular ROI tests first-order quantities under offsets $-7$, $0$,
and $9$. Translations of the ROI by $-2,-1,1,2$ voxels along $x$ have an analytic
mean change of twice the signed displacement. Expected values are calculated
directly from the object formula and ROI coordinates.

Paired deltas are checked for positive, negative, zero, and near-zero denominators.
Scalar reference calculations test the implemented two-way ICC variants~\cite{icc,mcgraw} and
Lin's concordance correlation coefficient (CCC) on complete synthetic pairs.
CCC separates association from location and scale agreement~\cite{lin}:
\begin{equation}
 \rho_c=\frac{2\sigma_{xy}}{\sigma_x^2+\sigma_y^2+(\mu_x-\mu_y)^2}
        =\rho C_b,\qquad
 C_b=\frac{2\sigma_x\sigma_y}{\sigma_x^2+\sigma_y^2+(\mu_x-\mu_y)^2}.
\label{eq:ccc}
\end{equation}
Equation~\ref{eq:ccc} permits a defined $C_b$ even when $\rho=0$, an edge case
included in the checks.

\subsection{Technical evaluation of the qualitative workflow}
The qualitative evaluation uses simulated responses only. Independent coefficient
references cover the applicable ordinal agreement and ranking calculations;
ordinal alpha follows its marginal-frequency distance definition~\cite{krippendorff}.
Chance agreement for the AC2 comparison follows the author's reference
implementation~\cite{gwet}. Balanced, skewed, incomplete, and degenerate fixtures
are included. Known seeds allow the resampling algorithm to be compared with
an independently coded calculation of the same resamples.
The software also exposes Cohen's unweighted and weighted kappa~\cite{cohen,weightedkappa},
Fleiss' kappa~\cite{fleiss}, and ICC~\cite{icc,mcgraw}. Their presence in the software
is documented separately from the ordinal-alpha/AC2 phantom fixtures reported here.
Ranking agreement uses Kendall's $W$~\cite{kendall}; applicable paired comparisons
and independent-sample probability-of-superiority summaries follow Wilcoxon and
Mann--Whitney methods~\cite{wilcoxon,mannwhitney}, with Holm adjustment where a
family of comparisons is requested~\cite{holm}. Independent-sample summaries do
not account for crossed reader/case dependence; the crossed bootstrap below
addresses the paired reader-by-case comparison.

Actual synthetic image packages are authored, converted to blinded evaluator
deliveries, supplied with simulated responses, reopened, deblinded, and aggregated.
Checks concern the saved order, hidden repeats, public/private metadata separation,
source-role recovery, and missing-response denominators. Sentinel metadata embedded
in synthetic files tests whether non-pixel information survives delivery.

The master seed was 20260909. Nine coefficient fixtures and six ranking fixtures
covered complete, incomplete, skewed, tied, and degenerate responses. Twelve bootstrap
endpoint checks compared identical seeded resamples with independent scalar calculations.
Twenty-four presentation plans each used 12 pages, three panels, three questions, and
four hidden repeats with at least two intervening pages. Four complete evaluator
deliveries supplied 127 simulated answer records.

A separate missing-response fixture supplied four evaluator assignments:
one answered, one skipped, one not assessable, and one entirely absent response.
An assignment roster identifies entirely absent responses; without a roster,
the denominator contains only received records.

For interval calibration, differences followed
$D_{rc}=0.5+A_r+B_c+\epsilon_{rc}$, with independent zero-mean Gaussian reader, case,
and residual effects of standard deviations 0.35, 0.50, and 0.40. Each of two settings
(complete data and 15\% missing completely at random) used 100 independently generated
datasets, four readers, 12 cases, and 199 independently sampled reader-and-case bootstrap
replicates, following separate resampling of crossed rows and columns~\cite{owen}.
Coverage is the fraction of intervals containing the population median difference 0.5.

Binomial uncertainty in the Monte Carlo coverage estimates is summarized using
Wilson score intervals~\cite{wilson}, separately from the bootstrap intervals
whose coverage is being assessed.

\clearpage
\subsection{Workbench views}
\label{app:gui}
The blinded evaluator view and paired radiomics workspace are shown in
Figures~\ref{fig:gui-qualitative} and \ref{fig:gui-radiomics}. Captures use the
saved synthetic filtering comparison and a monochrome publication palette.\par\medskip
\noindent\begin{minipage}{\linewidth}\centering\captionsetup{hypcap=false}
\includegraphics[width=\linewidth]{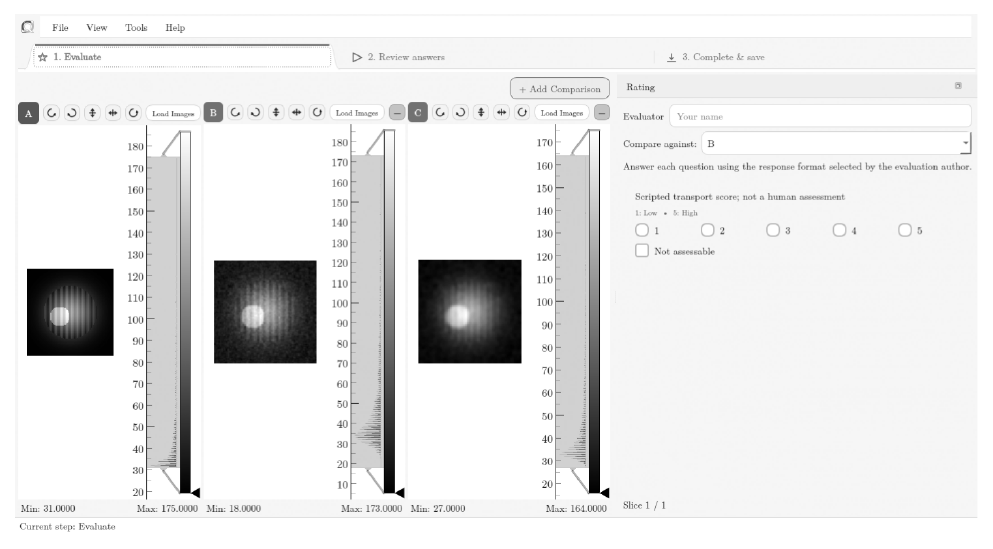}
\captionof{figure}{Blinded evaluator view opened from a delivered package. The presentation
labels and response form come from the exported study.}
\label{fig:gui-qualitative}
\end{minipage}\par\medskip
\noindent\begin{minipage}{\linewidth}\centering\captionsetup{hypcap=false}
\includegraphics[width=\linewidth]{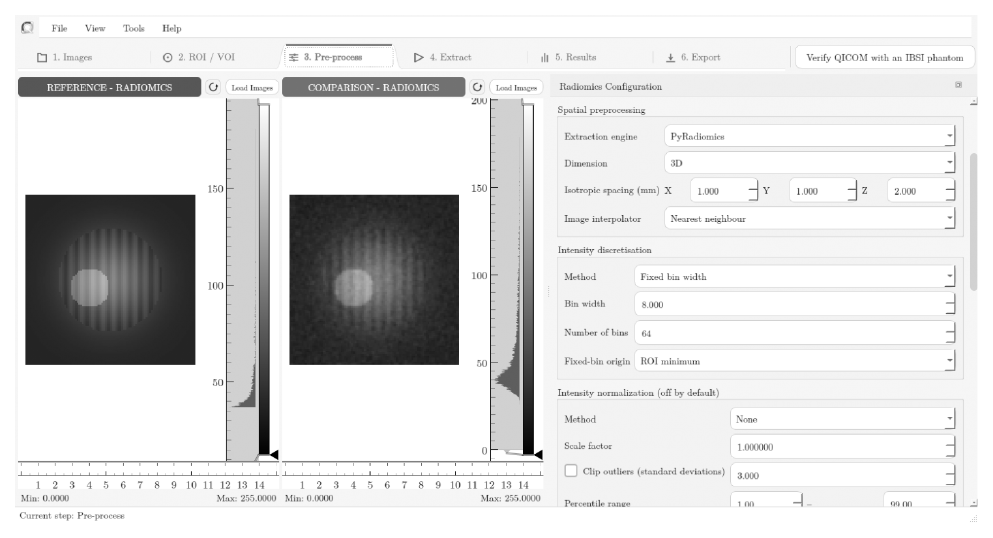}
\captionof{figure}{Paired radiomics view reopened from a native project, showing the images and
extraction settings used in the filtering comparison.}
\label{fig:gui-radiomics}
\end{minipage}\par\medskip

\clearpage
\section{Reproduction records}
\label{app:reproduction}
The companion bundle contains two source archives with file-level SHA-256
manifests. The original archive accompanies the quantitative, diagnostic,
qualitative, PyRadiomics, and synthetic filtering experiments. The extension
archive accompanies the complete IBSI~1 evaluation and the GUI captures.
Both use QICOM version 1.1.0.dev0 and Python 3.12.3 on Linux x86\_64.
\begin{description}
\item[Original application SHA-256] \hfill\\{\small\url{b1c3cdbdfa96140f9822f8ad8542e0a9fba24a41e44a8fe25ab695afab34a922}}
\item[Extended application SHA-256] \hfill\\{\small\url{808376b3cc766a6228f316cff1d4fd354c72a2e6294d796e0e88beaf6ebdf3f7}}
\end{description}
\begin{table}[htbp]
\centering\small
\begin{tabular}{ll}
\toprule
Computation package & Installed version \\
\midrule
lpips & \texttt{0.1.4} \\
nibabel & \texttt{5.4.2} \\
numpy & \texttt{2.5.2} \\
piq & \texttt{0.8.0} \\
scikit-image & \texttt{0.26.0} \\
scipy & \texttt{1.18.1} \\
torch & \texttt{2.14.0+cpu} \\
torchvision & \texttt{0.29.0+cpu} \\
mirp & \texttt{2.7.0} \\
itk & \texttt{5.4.7} \\
SimpleITK & \texttt{2.5.6} \\
\bottomrule
\end{tabular}
\caption{Principal computation dependencies. The complete environment manifests include additional packages and learned-weight identities.}
\label{tab:versions}
\end{table}

PyRadiomics is pinned to source commit
\begin{quote}\small\url{6a761c4eb01905cfcb1139b0b158547c55edd4df}\end{quote}
Its runtime label is \texttt{3.1.0a2.post8+g6a761c4} and distribution-metadata
label is \texttt{3.0.1a1}; both identify the same installation. The extension uses
MIRP~2.7.0. Receipt files identify the backend for every radiomic comparison.

The experiment folders contain executable configurations, inputs, numerical
references, observations, and full comparison tables. The bundle README provides
the reproduction commands. Failed development runs and regression logs are
retained as companion records. Official IBSI resources and learned checkpoints
are obtained separately from the recorded URLs and verified hashes. Analyses
run offline once these resources are present. The LaTeX sources include the
compiled bibliography and figures; compilation used Tectonic~0.17.0.

\clearpage
\section{Software attribution}
\label{app:software}
Numerical arrays and algorithms use NumPy~\cite{numpy}, SciPy~\cite{scipy}, and
scikit-image~\cite{skimage}; result tables use pandas~\cite{pandas,pandas_software}.
DICOM and NIfTI I/O use pydicom~\cite{pydicom} and NiBabel~\cite{nibabel}, respectively.
SimpleITK supports physical-space image operations~\cite{simpleitk}, while
PyRadiomics provides feature extraction~\cite{pyradiomics}, including optional
wavelet filtering through PyWavelets~\cite{pywavelets}. The IBSI digital-phantom
experiment uses unfiltered images. The optional MIRP backend~\cite{mirp}, with
ITK~\cite{itk}, supplies the extended IBSI~1 extraction. QICOM computes the exact
spatial autocorrelations and preprocessing diagnostics described in the Methods.
Learned and perceptual backends use
PyTorch~\cite{torch,torch2}, TorchVision~\cite{torchvision}, PIQ~\cite{piq,piqsoftware},
and LPIPS~\cite{lpips}. PyTorch supplies the execution framework. Figures were generated with
Matplotlib~\cite{matplotlib}. Exact installed versions are retained in the artifact.

The following supporting packages
provide the interface, storage, and execution infrastructure. Project references
are used when no preferred research article was found in the reviewed official
documentation. The companion citation audit records the sources consulted and
distinguishes requested scholarly citations from software license notices.

\begin{longtable}{>{\raggedright\arraybackslash}p{0.39\linewidth}>{\raggedright\arraybackslash}p{0.55\linewidth}}
\toprule
Package & Role in QICOM \\
\midrule
\endhead
Qt for Python / PySide6~\cite{pyside} & Desktop widgets, rendering, and worker orchestration \\
PyQtGraph~\cite{pyqtgraph} & Interactive image views, plots, and ROI interaction \\
Pillow~\cite{pillow} & Raster-image reading, writing, and blinded-delivery metadata handling \\
dicom-anonymizer~\cite{dicom_anonymizer} & DICOM de-identification support within QICOM's additional checks \\
openpyxl~\cite{openpyxl} & Spreadsheet report generation \\
PyYAML~\cite{pyyaml} & Metric descriptors and configuration serialization \\
platformdirs~\cite{platformdirs} & Platform-specific application data and configuration locations \\
threadpoolctl~\cite{threadpoolctl} & Numerical-library thread limits during parallel execution \\
cryptography~\cite{cryptography} & Authenticated encryption of evaluator files and access-code protection, in versions after the evaluated snapshots \\
GDCM / python-gdcm~\cite{gdcm} & Decoding of lossless JPEG and JPEG-LS DICOM pixel data, in versions after the evaluated snapshots \\
\bottomrule
\end{longtable}
Third-party license and copyright notices are included separately in the software
distribution. The dependency manifest identifies optional packages. 3D Slicer is
cited as a related platform and irrCAC as a methodological reference.

\clearpage
\section{Later software version: workbenches and Utility Tools}
\label{app:tools}
This section describes QICOM 1.1.0.dev0 at repository commit \texttt{dcdd7a8}
(19 October 2026), a development version later than the snapshots evaluated in this study
(Supplementary Section~\ref{app:reproduction}). Its
additions were not used in the experiments, and no value visible in these captures is
a result of the study. The captures come from the running application, rendered from
its Qt widgets with the monochrome interface palette of Supplementary
Section~\ref{app:gui}; label and mask overlays keep their colours.

All captures use one synthetic thorax phantom in Hounsfield units (HU): 24 axial slices
of $256\times256$ pixels, with $0.8\times0.8$~mm pixels and 2.5~mm slice spacing. It
contains air, a patient table, subcutaneous fat, soft tissue, two lungs, the heart,
a contrast-filled aorta, a vertebra with cortical and trabecular bone, and ribs, all
drawn as analytic shapes. Two DICOM series imitate a smooth and a sharp
reconstruction through in-plane Gaussian smoothing (standard deviation 1.3 and 0.4
pixels) and seeded Gaussian noise (8 and 24~HU). They carry the kernel labels B30f and
B70f only as header values: no reconstruction was simulated. The capture script, the
phantom generator and a manifest with the voxel hashes of both series are kept in
\texttt{paper/utility-tools}.

\subsection{The three workbenches in the later version}
\label{app:later-workbenches}
The launcher (Figure~\ref{fig:later-launcher}) opens the three workflows and, as a
fourth section, the Utility Tools. Each workflow window shows its steps in a rail
along the top. The quantitative workbench has eight steps: Import, Inspect,
Measure, Remove air, Analyze, Results, Heatmaps and Export. Figure~\ref{fig:later-quantitative-menu}
shows the Inspect step with both phantom series in a soft-tissue window (level 40~HU,
width 400~HU), and the Tools menu open on the Utility Tools submenu. Seven
full-reference metrics computed with a data range of 4095 appear in the Results
step (Figure~\ref{fig:later-quantitative-results}). These values are specific to the
phantom pair.

\noindent\begin{minipage}{\linewidth}\centering\captionsetup{hypcap=false}
\includegraphics[width=\linewidth]{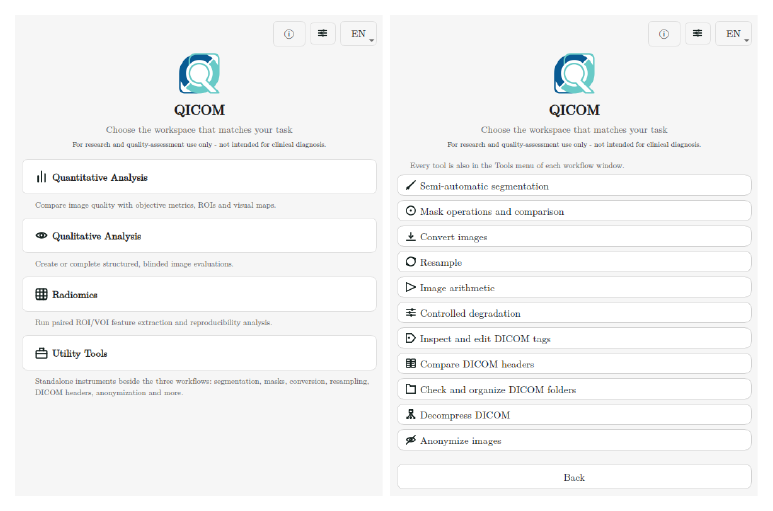}
\captionof{figure}{Launcher of the later version: the three workflows and
the Utility Tools section (left), and the tool catalog it opens (right).}
\label{fig:later-launcher}
\end{minipage}\par\medskip

\noindent\begin{minipage}{\linewidth}\centering\captionsetup{hypcap=false}
\includegraphics[width=\linewidth]{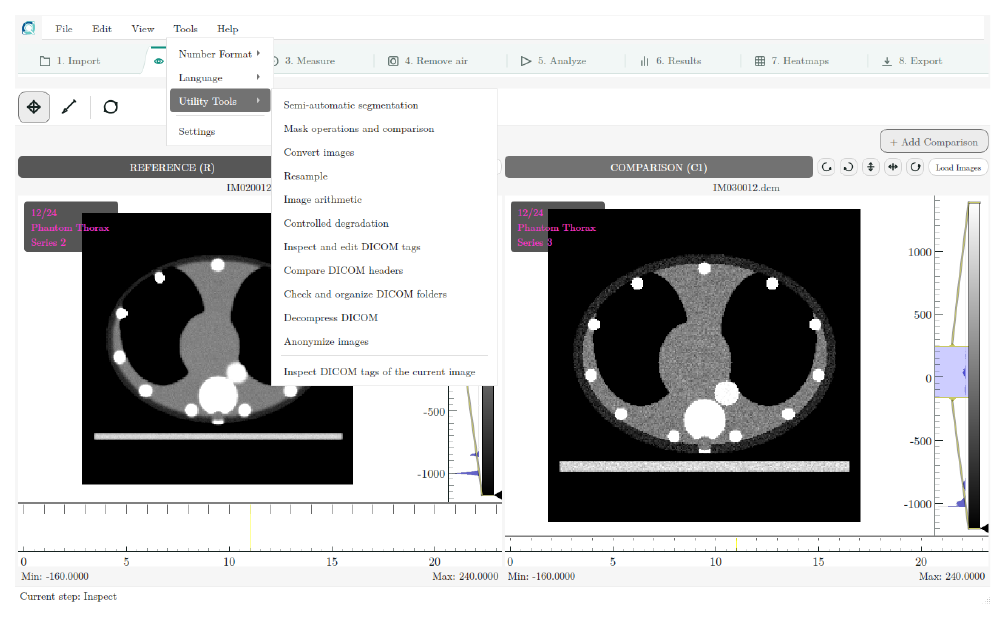}
\captionof{figure}{Quantitative workbench, Inspect step, with the
two phantom series and the Utility Tools submenu of the Tools menu. The two menus are
rendered separately and placed where Qt opens them.}
\label{fig:later-quantitative-menu}
\end{minipage}\par\medskip

\noindent\begin{minipage}{\linewidth}\centering\captionsetup{hypcap=false}
\includegraphics[width=\linewidth]{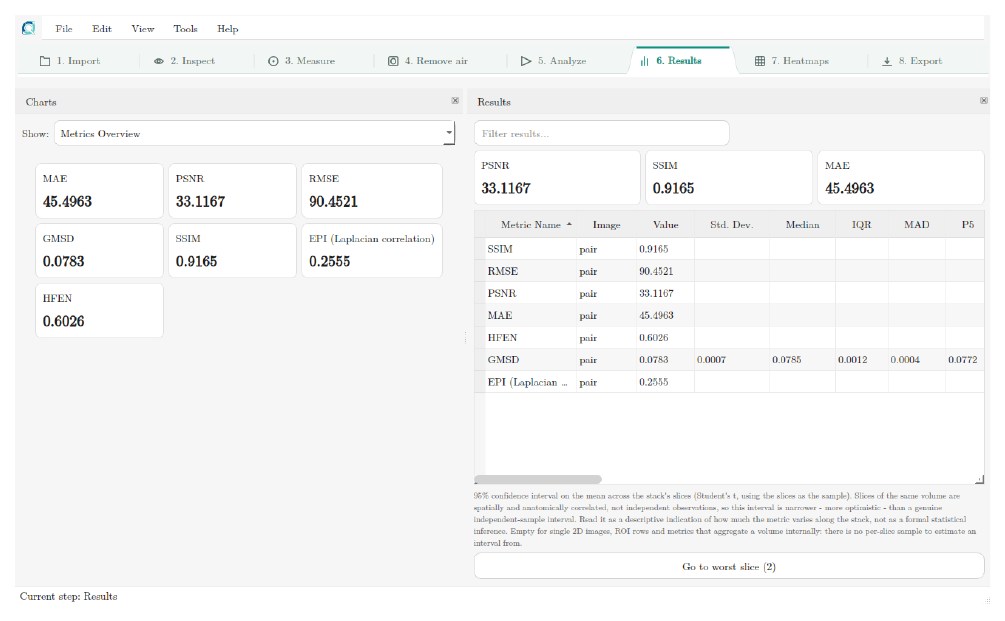}
\captionof{figure}{Quantitative workbench, Results step: the
sharp series compared with the smooth reference.}
\label{fig:later-quantitative-results}
\end{minipage}\par\medskip

The qualitative workbench separates the author's steps (Setup, Images, Questions, All
pages, Preview and Save) from the evaluator's (Evaluate, Review answers, and
Complete and save). Figure~\ref{fig:later-qualitative-author} shows the Images step of a
page with two questions. The author has named the panels after the two series. The
blinded delivery made from it (Figure~\ref{fig:later-qualitative-evaluator}) shows the
same images under neutral labels in a randomized order, with the response form beside
them; this evaluator is not logged in, so no answer is kept on the computer. For images
with several slices the author chooses whether answers apply to the whole series, the
default, or to each slice separately. With the first, one answer is kept per page,
comparison and question, and the evaluator scrolls through the slices without losing it;
the slice indicator below the form states which rule applies. The randomization seed, the
master seed and the hidden repeats are kept only in the author's private map: the
evaluator's package numbers pages and questions in the order presented, so it does not
reveal which panel is the reference or which pages are repeats. Every image is
de-identified again on its way into a delivery. A refusal stops the delivery before any
file is written, and the author decides whether to include those images in that delivery
alone.

\noindent\begin{minipage}{\linewidth}\centering\captionsetup{hypcap=false}
\includegraphics[width=\linewidth]{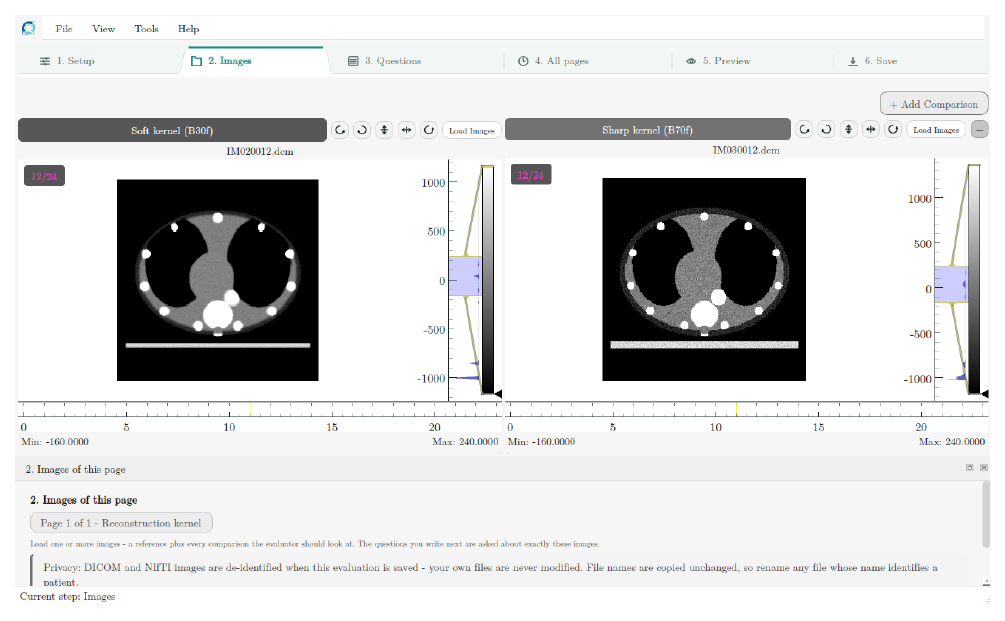}
\captionof{figure}{Qualitative workbench, author side: the Images
step of a page built on the two phantom series.}
\label{fig:later-qualitative-author}
\end{minipage}\par\medskip

\noindent\begin{minipage}{\linewidth}\centering\captionsetup{hypcap=false}
\includegraphics[width=\linewidth]{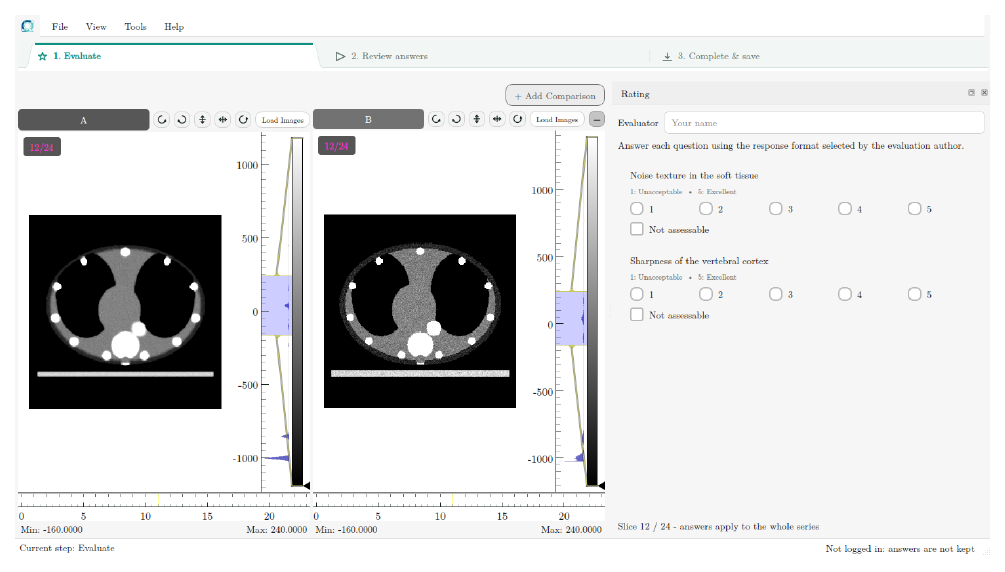}
\captionof{figure}{Qualitative workbench, evaluator side: the
blinded delivery, with neutral panel labels and the two questions.}
\label{fig:later-qualitative-evaluator}
\end{minipage}\par\medskip

The radiomics workbench has six steps: Images, ROI/VOI, Pre-process, Extract, Results
and Export. The project shown was computed with QICOM's radiomics functions and
reopened in the workbench: the lung label of the segmentation in
Figure~\ref{fig:tool-segmenter-labels} as the VOI, linear interpolation to the native
$0.8\times0.8\times2.5$~mm spacing, and the PyRadiomics backend with QICOM's default feature
classes and a bin width of 25~HU. Figure~\ref{fig:later-radiomics-roi} shows the VOI on both series in a
lung window (level $-600$~HU, width 1500~HU), and Figure~\ref{fig:later-radiomics-results}
the delta-radiomics table. A project's feature tables are saved both as CSV, for other
software, and as typed JSON, which QICOM reads back value by value, so that a reopened
table still matches the hash recorded in its receipt.

\noindent\begin{minipage}{\linewidth}\centering\captionsetup{hypcap=false}
\includegraphics[width=\linewidth]{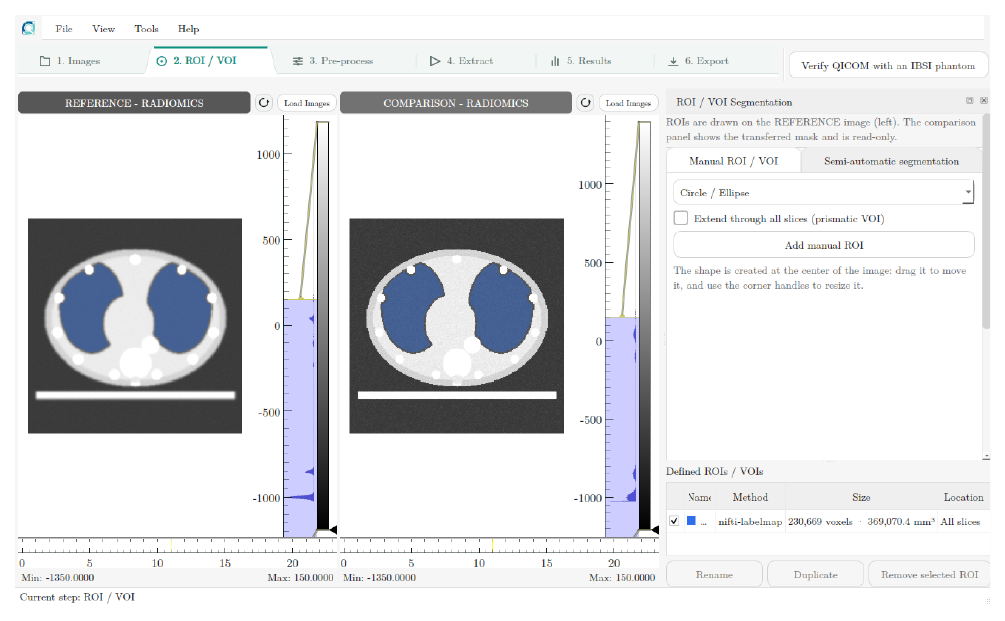}
\captionof{figure}{Radiomics workbench, ROI/VOI step: the lung label
from the segmenter on both phantom series.}
\label{fig:later-radiomics-roi}
\end{minipage}\par\medskip

\noindent\begin{minipage}{\linewidth}\centering\captionsetup{hypcap=false}
\includegraphics[width=\linewidth]{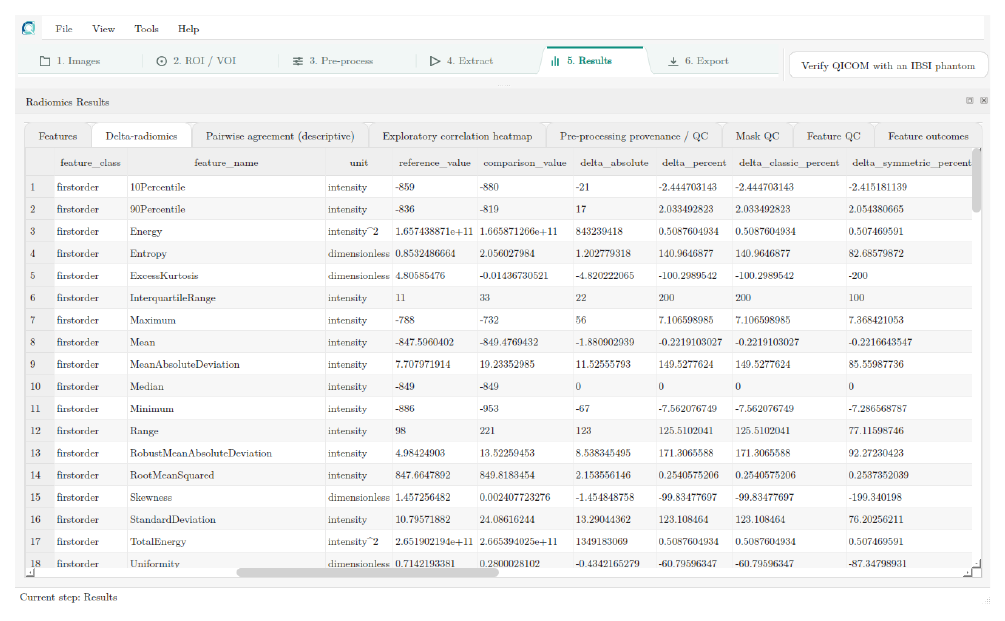}
\captionof{figure}{Radiomics workbench, Results step: reference
and comparison values and their differences for the lung VOI.}
\label{fig:later-radiomics-results}
\end{minipage}\par\medskip

\clearpage
\subsection{Utility Tools}
\label{app:utility-tools}
The Utility Tools are eleven self-contained tools in one window, with the catalog on the
left and the selected tool on the right. Opened from the launcher, the window returns
to it when closed. Each workflow window can also open its own copy from the Tools menu.
That copy is reused while the workflow window is open and closes with it. The
quantitative and radiomics windows also offer the DICOM header of the image they have
loaded. The qualitative window does not, so an evaluator answering a blinded package is
not offered the header of the image being rated.

The tools share four conventions. Sources are only read. Every set of output files is
built in a hidden staging folder and renamed into a new or empty destination at the
end, so it appears complete or not at all; the destination may be neither a link nor
the source folder. A NIfTI file written for a DICOM source is placed by the source's
position, orientation and spacing, with the column, row and slice layout of dcm2niix
and SimpleITK; the tests compare it with SimpleITK on an oblique series. Slices are
placed by the positions the loader delivers in pixel order. A stack that no single
affine describes, with irregular steps, repeated positions or frames in different
orientations, is written with qform and sform codes 0, which declare its geometry
unknown, and a sheared stack keeps its shear in the sform, which the qform cannot
represent. Finally, no DICOM or NIfTI header field or text chunk of a source is copied
into a derived image.

DICOM pixel data in lossless JPEG or JPEG-LS, the encoding of most CT and MRI exports
from scanners and archives, are decoded through GDCM~\cite{gdcm}. Baseline JPEG and
JPEG~2000 remain decoded by Pillow, so the pixels of files that QICOM already read do
not change. This applies to the workflows as well as to the tools.

The tools also handle alike. Every computation and every save runs outside the
interface thread and shows its progress as a percentage in a window of its own. In the
image area, as in the workflows, the mouse wheel steps through the slices and Ctrl and
the wheel zoom. A trackpad gesture is summed into whole wheel notches and its inertia is
ignored. A left drag sets window and level,
as in common DICOM viewers: horizontally the window width, which a drag to the right
widens, and vertically the level, which a downward drag raises, darkening the image. The
middle button pans and the right button zooms. While nothing is loaded, a click on the
area, or a file dropped on it, opens an image. On every tool one kind of button is coloured at a time: the step to
take next, as in the air removal of the quantitative workbench.

\subsubsection{Semi-automatic segmentation}
\emph{What it does.} It builds a label map one sampled region at a time. Labels are
named by the user or taken from presets (air, lung, fat, soft tissue, contrast medium,
bone, metal and patient table), and a ticked list selects which labels are shown.

\emph{What it is for.} It separates intensity classes: removing air and the table
before a comparison, isolating bone, or preparing a VOI. A saved label map is imported
by the radiomics workbench on the same voxels.

\emph{How it works.} The tool is a loop of six pages shown as tabs: Image, Label,
Sample, Selection, Apply and Save, with Back and Next below them. Each pass picks a
label, samples the pixels under a rectangular, elliptical or freehand ROI chosen on the
Sample page, and turns them into an intensity range: percentiles
(1--99 by default) or mean $\pm k$ standard deviations. A preset can instead start from
an indicative CT range, adjusted on the histogram. The user chooses the reach of the
selection, the current slice or the volume, anywhere in range or only connected to
the ROI, and optional refinements: median smoothing, hole filling, removal of small
regions, opening and closing. The selection is previewed before anything changes. It
is then removed from the image, kept alone, assigned to the label, or erased from it,
and "Segment another region" leads back to choosing the next label.
Each pixel holds at most one label, and a step takes only unassigned pixels unless
told to overwrite, so the order of steps resolves overlapping ranges. Every step can be
undone. The map is saved as an 8-bit NIfTI placed on the source grid, with a versioned
JSON sidecar holding label names, colours and steps but no source file name, path or
header value. The method separates intensity ranges, not organs: liver, spleen and
muscle share a CT range, and the preset ranges are starting points, not validated
thresholds.

\noindent\begin{minipage}{\linewidth}\centering\captionsetup{hypcap=false}
\includegraphics[width=\linewidth]{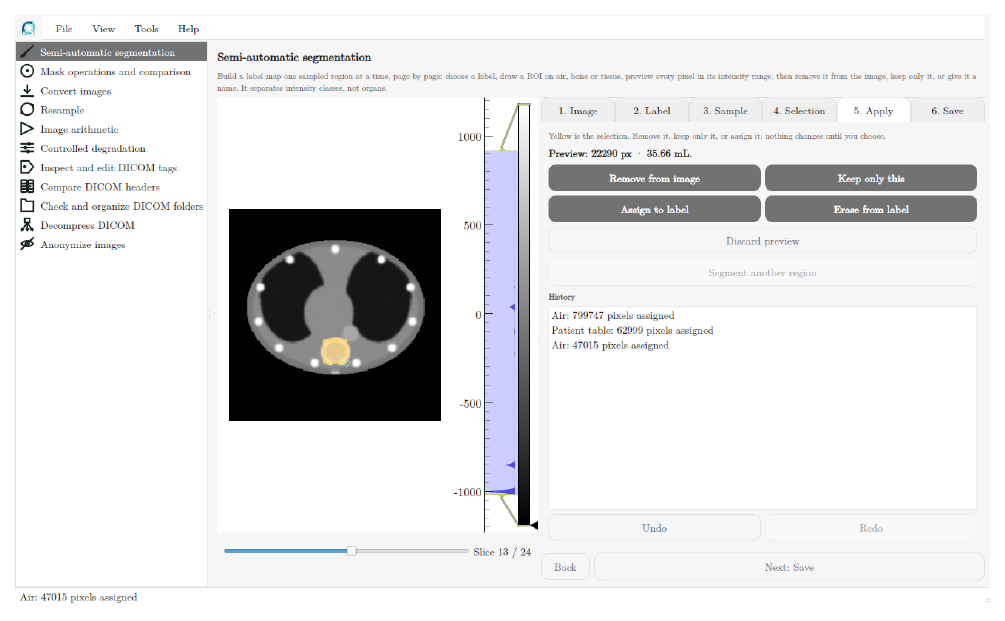}
\captionof{figure}{Segmenter, Apply page: after air and the table were
removed, the bone preset range is previewed in yellow, connected to the ROI on the
vertebra, and the four actions are the step due.}
\label{fig:tool-segmenter-step}
\end{minipage}\par\medskip

\noindent\begin{minipage}{\linewidth}\centering\captionsetup{hypcap=false}
\includegraphics[width=\linewidth]{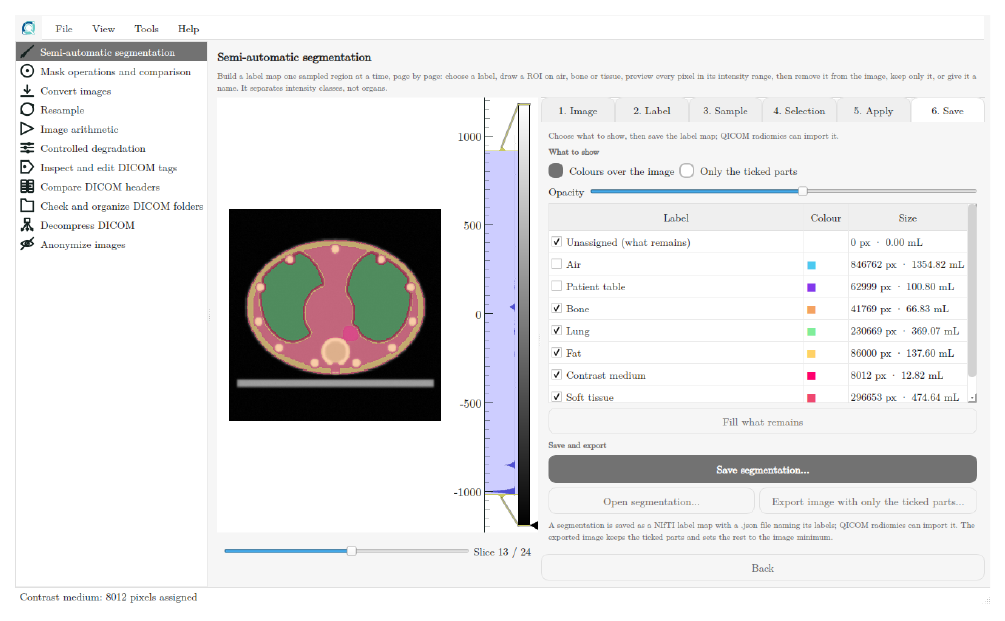}
\captionof{figure}{Segmenter, Save page with the complete label map: bone
(vertebra and ribs), lungs (each connected to its own ROI), fat, contrast medium, and the
remaining soft tissue, with air and table hidden.}
\label{fig:tool-segmenter-labels}
\end{minipage}\par\medskip

\subsubsection{Mask operations and comparison}
\emph{What it does.} It combines two masks by union, intersection, difference or
exclusive or. It also fills the empty slices between drawn ones, and compares two
masks by voxel counts, volumes, Dice and Jaccard coefficients, relative volume
difference, Hausdorff distance, its 95th percentile and mean surface distance.

\emph{What it is for.} It compares segmentations from two readers or methods, merges
labels, and completes contours drawn on a few slices.

\emph{How it works.} Each mask is a NIfTI label map read as any non-zero value or as
one chosen label, so a multi-label map can be compared structure by structure. Two
masks must share a grid: nothing is resampled for a comparison. Slice filling
interpolates signed distance maps between drawn slices. Distances are in millimetres
from each file's voxel spacing, and a surface distance is undefined when either mask
is empty. The overlay colours voxels in A only, in B only, in both, and in the last
result; the comparison table can be copied as CSV.

\noindent\begin{minipage}{\linewidth}\centering\captionsetup{hypcap=false}
\includegraphics[width=\linewidth]{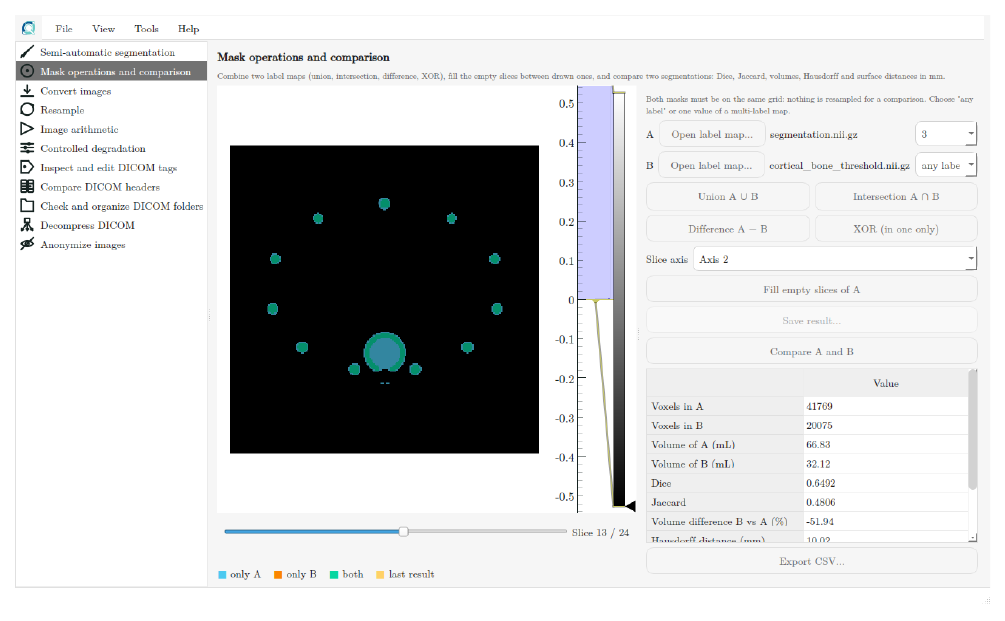}
\captionof{figure}{Mask comparison: the segmenter's bone label (A) against a
threshold above 700~HU, which keeps cortical bone only (B).}
\label{fig:tool-masks}
\end{minipage}\par\medskip

\subsubsection{Image conversion}
\emph{What it does.} It converts a DICOM file or series, a NIfTI volume or a bitmap stack
to a NIfTI volume, to 8- or 16-bit PNG slices, or to 32-bit floating-point TIFF slices.

\emph{What it is for.} It prepares inputs for other software and images for figures.

\emph{How it works.} A NIfTI volume is placed by the source geometry and stores whole
numbers in the smallest integer type that represents every value exactly, so CT in HU
fits in 16 bits; other values are stored as 32-bit floats. A bitmap without geometry
receives unit spacing and declares it unknown. PNG slices use the window on screen,
which suits figures but not measurement, with 256 or 65\,536 grey levels across it;
TIFF slices keep the physical values. The user picks a folder, and the tool creates a
new one inside it named after the source and the format
(\texttt{<name>\_png16}, numbered when the name is taken), so converted files never mix
with what the folder already holds.

\noindent\begin{minipage}{\linewidth}\centering\captionsetup{hypcap=false}
\includegraphics[width=\linewidth]{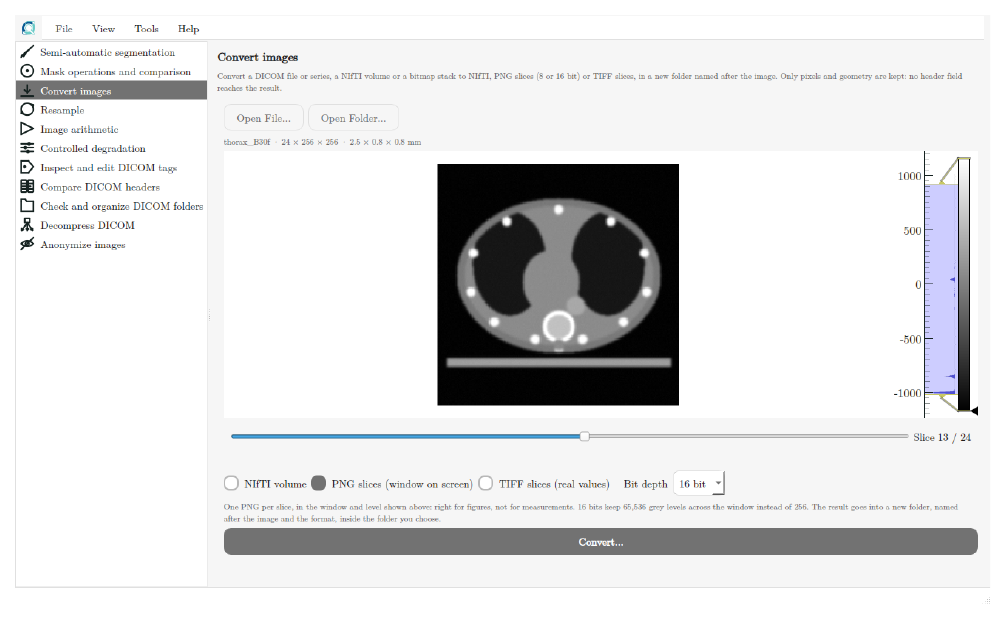}
\captionof{figure}{Image conversion of the smooth series to 16-bit PNG slices
in the window shown.}
\label{fig:tool-converter}
\end{minipage}\par\medskip

\subsubsection{Resampling}
\emph{What it does.} It puts an image on another grid: a chosen voxel spacing, such as
isotropic voxels, or the grid of a reference image.

\emph{What it is for.} QICOM's quantitative comparison is voxel by voxel, so two images
must first share a grid.

\emph{How it works.} Resampling to a spacing keeps the position of the first voxel and
changes the step between voxels. Resampling onto a reference uses both images'
voxel-to-patient mappings and therefore assumes a shared frame of reference: images
are placed by their coordinates, not registered. Interpolation is nearest neighbour,
which keeps label values, linear or cubic, with a chosen value outside the image.
Nearest and linear interpolation are computed a block of output slices at a time,
which is what their progress counts.

\noindent\begin{minipage}{\linewidth}\centering\captionsetup{hypcap=false}
\includegraphics[width=\linewidth]{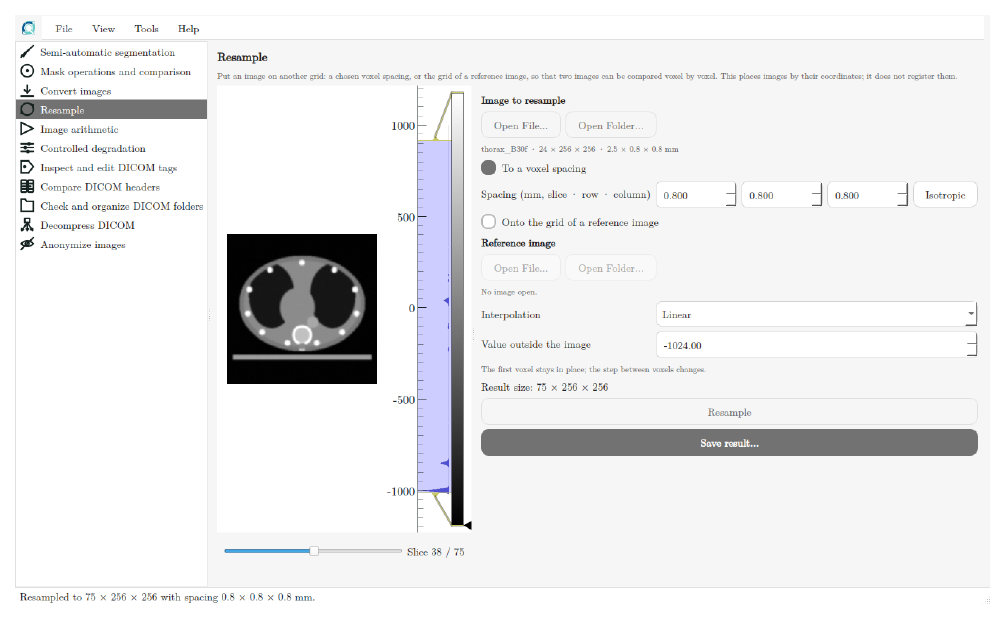}
\captionof{figure}{Resampling of the smooth series to isotropic 0.8~mm
voxels, which gives 75 slices.}
\label{fig:tool-resampling}
\end{minipage}\par\medskip

\subsubsection{Image arithmetic}
\emph{What it does.} It computes the difference, absolute difference, sum, mean or
ratio of images.

\emph{What it is for.} A difference image shows where two reconstructions disagree; the
mean of repeated acquisitions lowers their noise; a ratio shows relative change.

\emph{How it works.} Operations are voxel by voxel. The images must have the same size
and spacing, and the result is a 32-bit float image in the images' units. Zero
denominators of a ratio are counted and replaced by a chosen value instead of
infinities.

\noindent\begin{minipage}{\linewidth}\centering\captionsetup{hypcap=false}
\includegraphics[width=\linewidth]{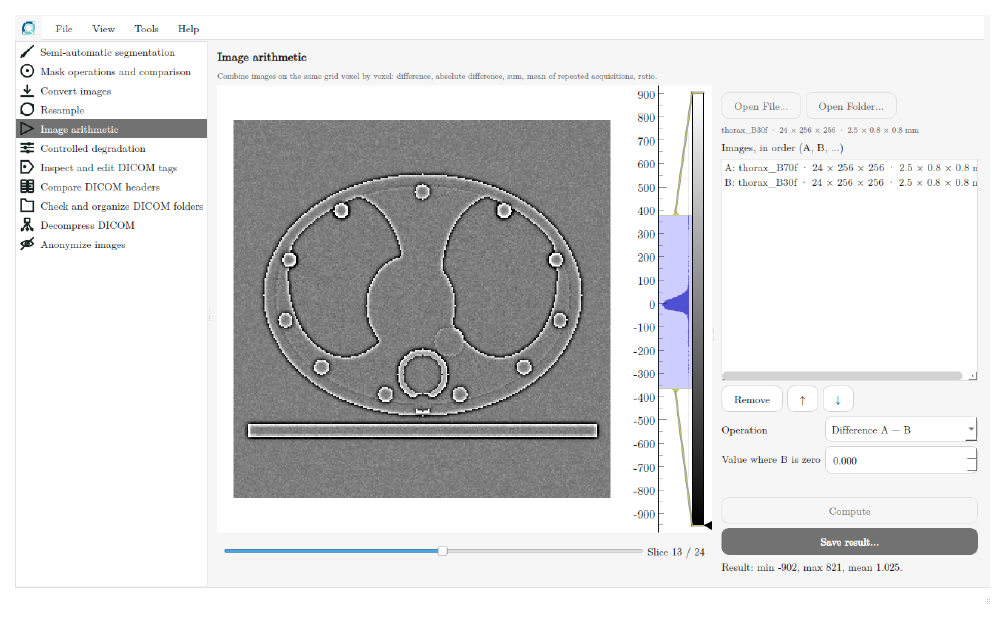}
\captionof{figure}{Image arithmetic: the sharp series minus the smooth
series.}
\label{fig:tool-arithmetic}
\end{minipage}\par\medskip

\subsubsection{Controlled degradation}
\emph{What it does.} It applies Gaussian noise, Poisson noise, Gaussian blur, loss of
resolution and JPEG compression, one effect at a time.

\emph{What it is for.} It produces test inputs to check that a metric responds in the
expected direction, for example more noise giving a worse score.

\emph{How it works.} The right column reads in the order of the work: one button per
effect, the strength of the chosen effect, Apply, the list of effects applied (each
removable), and saving. Every change recomputes the whole list from the original
image. All random draws come from one generator seeded by the user, so the
same request reproduces the same image. Values stay in the image's units, and spatial
operations act within each slice. JPEG compresses an 8-bit window of each slice and
maps it back, which also clips values outside the window; the window is therefore a
recorded parameter. A versioned JSON record holds the steps, the seed and the SHA-256
of the source and result pixels.

\noindent\begin{minipage}{\linewidth}\centering\captionsetup{hypcap=false}
\includegraphics[width=\linewidth]{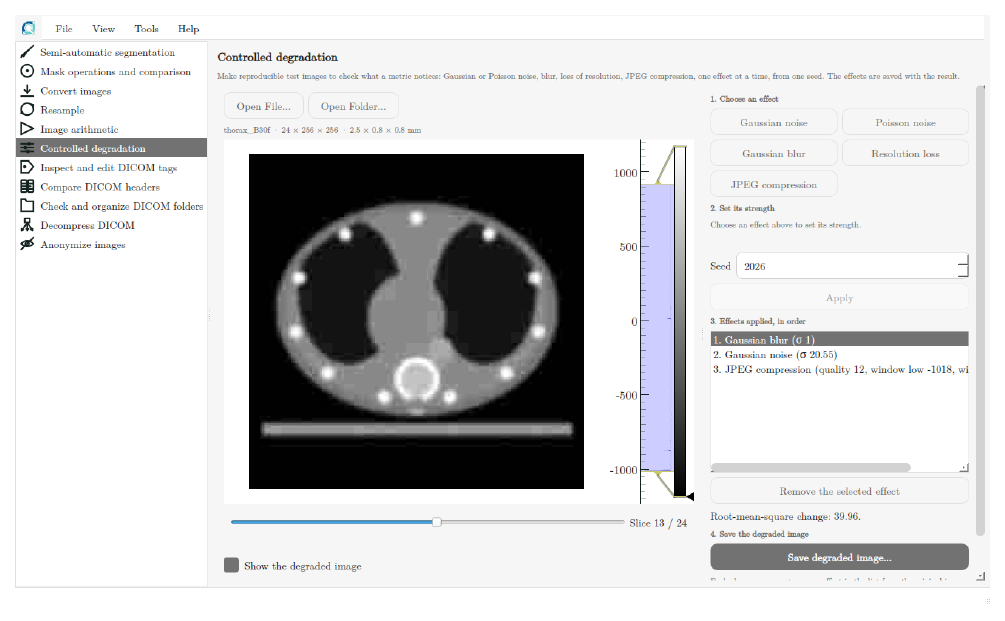}
\captionof{figure}{Controlled degradation of the smooth series: Gaussian
blur, Gaussian noise and JPEG compression at quality 12, applied one at a time.}
\label{fig:tool-degradation}
\end{minipage}\par\medskip

\subsubsection{DICOM tag inspection and editing}
\emph{What it does.} It lists every element of a DICOM file, or of one file of a series -
file meta information, nested sequences and private tags - with search, a filter for
private tags, copy and CSV export. In the same tree, a column takes a new value for
each top-level element that may be changed, added or deleted, in one file or in every
file of the series, and the result is written as copies in a new folder.

\emph{What it is for.} The cause of a difference between two images is often in the
header, such as a reconstruction kernel, slice thickness or tube current; and research
copies sometimes need correcting, for example a wrong series description, a missing
kernel name, or a pixel spacing lost in conversion.

\emph{How it works.} Reading stops before the pixel data. Sequences are flattened into
rows that keep their depth and path. A header is treated as untrusted input: listing
stops at 100\,000 rows and says so, and an element that cannot be decoded becomes a
single row marked unreadable. Values are shown as stored, patient data included, and
the tool states this. The originals are never modified. Elements that define how pixels
are stored, such as Rows, Bits Allocated or Photometric Interpretation, are locked, and
the file meta group is rebuilt rather than edited. Elements that change what a
measurement means, such as Pixel Spacing, Rescale Slope and Intercept, Slice Thickness
or the slice position and orientation, are marked and need confirmation before
writing. New values are checked against their value representation. The UIDs of the
copies can be regenerated consistently at the series, instance or study level, or
kept. A versioned JSON log records each tag, the action, the new value and the SHA-256
of every source and copy, but never the old value, which may be the identity the edit
removed. Only top-level elements with a text or numeric value representation are
edited; sequences and nested elements are kept as they are.

\noindent\begin{minipage}{\linewidth}\centering\captionsetup{hypcap=false}
\includegraphics[width=\linewidth]{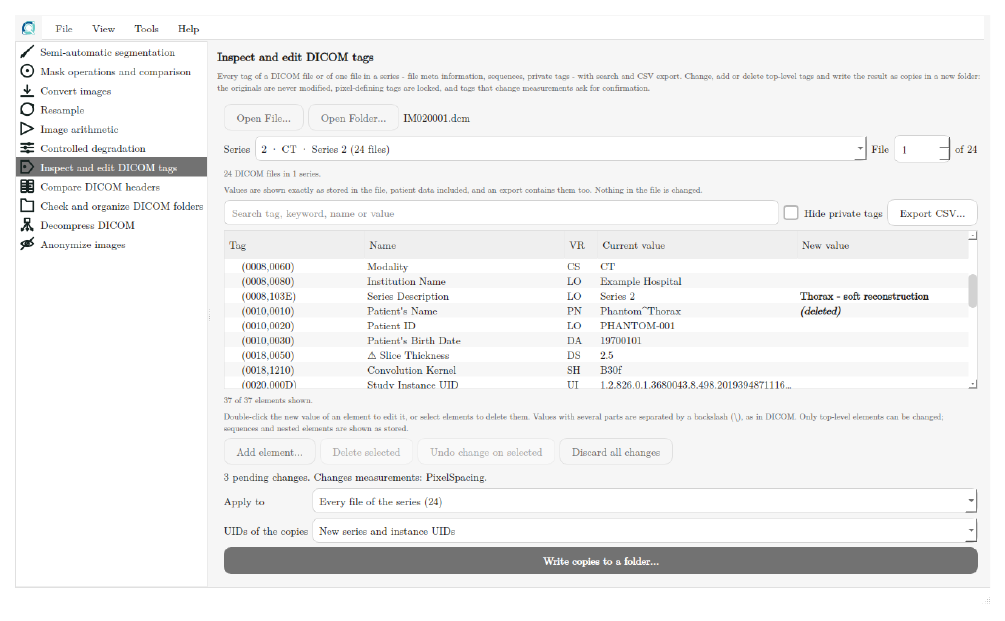}
\captionof{figure}{DICOM tag inspection and editing: the header of one file of the
smooth series, with three pending edits for the whole series, including the deleted
patient name and the flagged pixel spacing.}
\label{fig:tool-dicom-editor}
\end{minipage}\par\medskip

\subsubsection{DICOM header comparison}
\emph{What it does.} It shows two headers side by side, element by element, as
identical, different, or present in only one of them.

\emph{What it is for.} It explains why two acquisitions or reconstructions differ.

\emph{How it works.} Both headers are read as in the inspection and editing tool, and rows with the
same path are matched, so items of nested sequences are compared element by element.
The comparison opens on both whole headers; filters show only differences and hide
UIDs or private tags.

\noindent\begin{minipage}{\linewidth}\centering\captionsetup{hypcap=false}
\includegraphics[width=\linewidth]{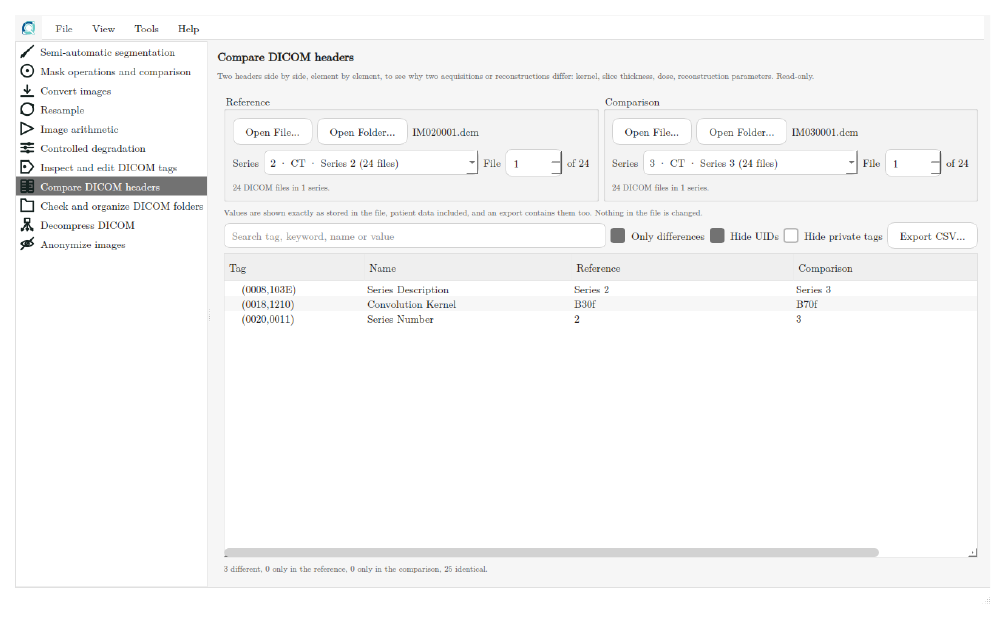}
\captionof{figure}{DICOM header comparison of the two phantom series,
showing only differences, with UIDs hidden.}
\label{fig:tool-dicom-compare}
\end{minipage}\par\medskip

\subsubsection{DICOM folder check and organization}
\emph{What it does.} It groups a folder of DICOM files by study and series and checks
each series for missing or repeated instance numbers, duplicate instances, slices of
different sizes, irregular slice spacing, mixed orientations, and slices without a
position. It exports the findings as CSV and can copy the files into an ordered tree.

\emph{What it is for.} Exports from an archive or a scanner often mix several series
in one folder; the check reports what would cause problems in a later comparison.

\emph{How it works.} Only headers are read, and links are not followed. The ordered copy
names folders by study order, series number and modality only
(\texttt{study\_01/series\_003\_CT/0001.dcm}), never by patient or description, but
keeps file contents unchanged; de-identification is the anonymizer's task.

\noindent\begin{minipage}{\linewidth}\centering\captionsetup{hypcap=false}
\includegraphics[width=\linewidth]{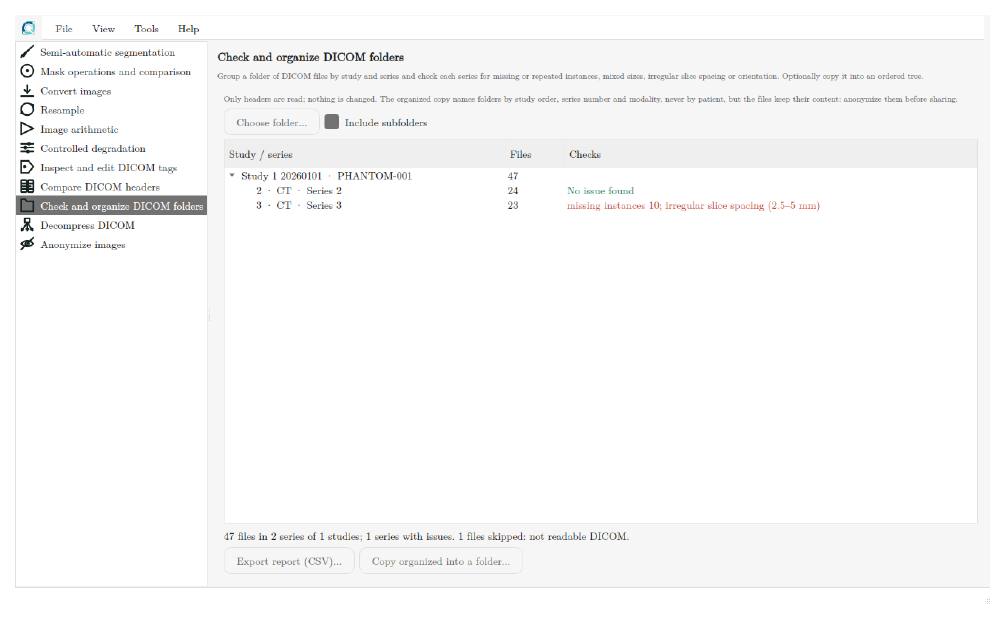}
\captionof{figure}{DICOM folder check of a folder that mixes both
phantom series, with one slice of the sharp series missing and a text file
skipped.}
\label{fig:tool-dicom-organizer}
\end{minipage}\par\medskip

\subsubsection{DICOM decompression}
\emph{What it does.} It rewrites a compressed DICOM file (lossless JPEG, JPEG-LS,
JPEG~2000, RLE or baseline JPEG) uncompressed, in Explicit VR Little Endian, one file per
frame.

\emph{What it is for.} Much of the software around QICOM cannot decode compressed pixel
data, and older readers also refuse multi-frame objects; the copies open in them.

\emph{How it works.} The header is read first, so before anything is written the tool
shows the encoding, the number of frames and of files to be written, and whether the
compression was lossy. The pixels are decoded by the same decoders, in the same order, as
QICOM's loader, so the copies hold exactly the pixels QICOM shows. The copies go into a
new folder named after the file, inside the folder the user picks, which starts as the
source's own folder, and appear there all together or not at all. A single-frame file
keeps every UID, since only its encoding changes. A multi-frame file becomes one instance
per frame: each copy receives a new SOP Instance UID, all of them one new Series Instance
UID, never mistaken for the source series, and the frame number as Instance Number, while
the study and the frame of reference stay. An Enhanced object stays Enhanced: each copy
carries its own per-frame functional group item and the shared ones, and vectors indexed
by the Frame Increment Pointer are cut to its frame. Decompression does not undo a lossy
compression: a copy of a baseline or extended JPEG source is marked Lossy Image
Compression ``01'' when the source did not say so.

\noindent\begin{minipage}{\linewidth}\centering\captionsetup{hypcap=false}
\includegraphics[width=\linewidth]{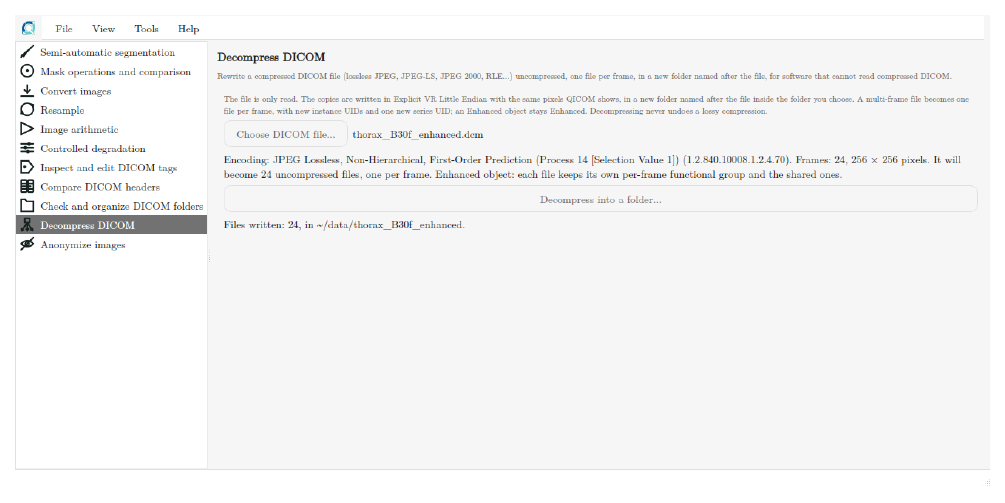}
\captionof{figure}{DICOM decompression: the smooth series stored as one Enhanced CT
object in lossless JPEG, 24 frames, rewritten as 24 uncompressed files.}
\label{fig:tool-dicom-decompressor}
\end{minipage}\par\medskip

\subsubsection{Image de-identification}
\emph{What it does.} It writes de-identified copies of DICOM, NIfTI and PNG, JPEG or BMP
files into a mirror of their folder.

\emph{What it is for.} It prepares images for sharing outside QICOM with the rules that
QICOM's own exports apply.

\emph{How it works.} The chosen files, or a folder with its subfolders, are first listed
as a tree, each with what will happen to it, together with the destination: a new folder
beside the original named like it with \texttt{\_anonymized} at the end, or the same
in another place the user picks. DICOM files receive the DICOM PS3.15 Basic Profile through
dicom-anonymizer on every nested sequence; private tags are removed, and UIDs, dates
and person names are replaced consistently across the set. The Part~10 container is
rebuilt rather than inherited: the 128-byte preamble is zeroed and the file meta group
keeps only the elements PS3.10 requires; each copy is read back and checked for leftover
identifying values before it is committed. A NIfTI file is recognized by its structure (magic string, dimensions,
data type, voxel offset and a complete payload), not by its first bytes; its header
text, extensions and gzip file names are cleared, everything between the header and the
voxels is zeroed, and bytes after the payload are dropped. Text metadata, EXIF and XMP are
removed from bitmaps. Burned-in text is not detected. A DICOM image is therefore refused unless it
declares that it carries no burned-in annotation, and refused when it declares
recognizable visual features; only a secondary-capture object without populated
identity attributes may omit the first declaration. A file that
cannot be de-identified stops the run before anything is written. The user may then
proceed explicitly, and those files are recorded as not verified. The copies keep the
subfolders and file names of the originals, so a de-identified export can stand in for
the original layout; the names themselves are not de-identified, which the tool states
before anything is written. For exports whose names identify the patient, an option
replaces them: folders become \texttt{folder\_001}, images \texttt{image\_0001.dcm},
in a folder named \texttt{anonymized}, and no table maps them back. Files that are not images are left out. The result should
still be checked against the institution's privacy procedure before it is shared.

\noindent\begin{minipage}{\linewidth}\centering\captionsetup{hypcap=false}
\includegraphics[width=\linewidth]{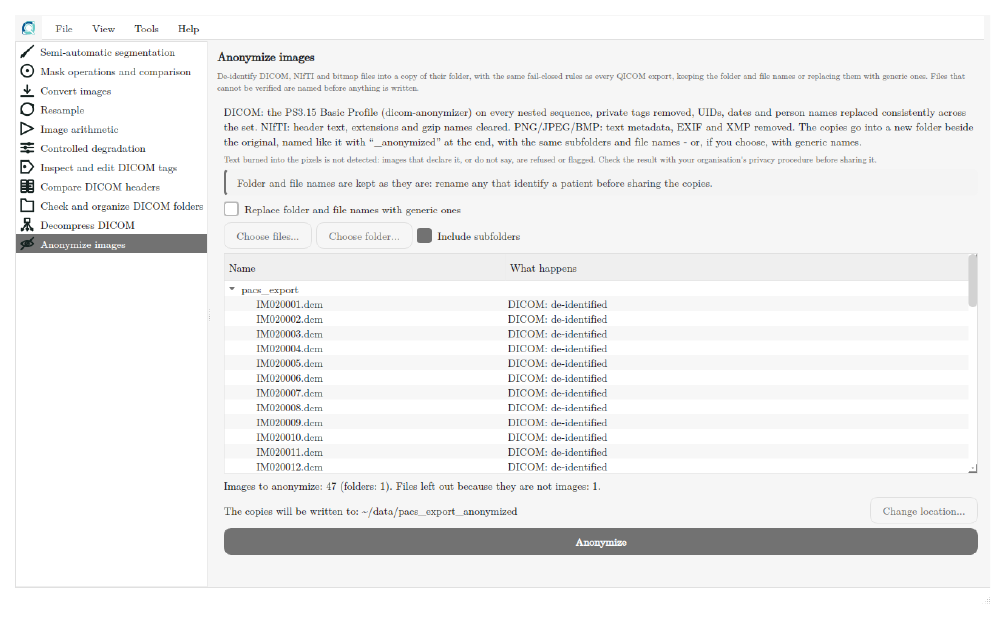}
\captionof{figure}{Image de-identification: the mixed folder listed as a tree, each
file with what will happen to it, and the destination \texttt{pacs\_export\_anonymized}
beside it.}
\label{fig:tool-anonymizer}
\end{minipage}\par\medskip

\end{document}